\documentclass[aps,prb,reprint,floatfix,superscriptaddress]{revtex4-2}
\usepackage{amsmath,amssymb}
\usepackage{graphicx}
\usepackage[bookmarks=false,colorlinks=true,linkcolor=blue,filecolor=blue,citecolor=blue,urlcolor=blue]{hyperref}
\usepackage{bm}

\begin{document}

\title{Multifractality in monitored single-particle dynamics}

\author{Kohei Yajima}
\affiliation{Department of Applied Physics, University of Tokyo, Tokyo 113-8656, Japan}

\author{Hisanori Oshima}
\affiliation{Department of Applied Physics, University of Tokyo, Tokyo 113-8656, Japan}
\affiliation{Nonequilibrium Quantum Statistical Mechanics RIKEN Hakubi Research Team, RIKEN CPR, Wako, Saitama 351-0198, Japan}

\author{Ken Mochizuki}
\affiliation{Department of Applied Physics, University of Tokyo, Tokyo 113-8656, Japan}
\affiliation{Nonequilibrium Quantum Statistical Mechanics RIKEN Hakubi Research Team, RIKEN CPR, Wako, Saitama 351-0198, Japan}

\author{Yohei Fuji}
\affiliation{Department of Applied Physics, University of Tokyo, Tokyo 113-8656, Japan}

\date{\today}

\begin{abstract}
We study multifractal properties in time evolution of a single particle subject to repeated measurements. 
For quantum systems, we consider circuit models consisting of local unitary gates and local projective measurements.
For classical systems, we consider models for estimating the trajectory of a particle evolved under local transition processes by partially measuring particle occupations. 
In both cases, multifractal behaviors appear in the ensemble of wave functions or probability distributions conditioned on measurement outcomes after a sufficiently long time.
While the nature of particle transport (diffusive or ballistic) qualitatively affects the multifractal properties, they are even quantitatively robust to the measurement rate or specific protocols. 
On the other hand, multifractality is generically lost by generalized measurements allowing erroneous outcomes or by postselection of the outcomes with no particle detection.
We demonstrate these properties by numerical simulations and also propose several simplified models, which allow us to analytically obtain multifractal properties in the monitored single-particle systems.
\end{abstract}

\maketitle
\tableofcontents

\section{Introduction}

Localization is a ubiquitous phenomenon in disordered systems, which leads to peculiar transport and thermalization properties. 
Free fermion systems subject to strong disorder can undergo the Anderson transition \cite{Evers08}, at which various physical measures exhibit universal, critical scaling behaviors. 
One of such measures is the moment of wave-function amplitude $\left| \psi (x) \right|^{2q}$, averaged over disorder. 
At the Anderson transition, it shows an algebraic scaling $L^{-\tau_q}$ with the system size $L$, which is governed by an exponent $\tau_q$ nonlinearly depending on $q$. 
This is understood as a consequence of \emph{multifractality} of the wave functions.
Multifractality at the Anderson transitions has been extensively studied in the context of metal-insulator transitions (between an extended and a localized phase) \cite{Wegner80, Castellani86, Janssen94, Grussbach95, Mirlin06, Subramaniam06, Vasquez08, Rodriguez08, Faez09, Obuse10, Rodriguez10, Rodriguez11, Gruzberg11, Gruzberg13, Padayasi23} or quantum Hall plateau transitions (between distinct localized phases) \cite{Ludwig94, Mudry96, Chamon96, Evers01, Evers08PRL, Obuse08, Zirnbauer19}. 
While multifractality in these systems only occurs at phase boundaries and thus requires fine-tuning of parameters in the systems, it can be extended to a finite region of the parameter space for long-range hopping models \cite{Mirlin96, Kravtsov97, Evers00, Mirlin00, Deng16, Nosov19}, models with correlated disorder \cite{Duthie22}, and quasiperiodic systems \cite{JunWang16, YuchengWang16, Roy18, Cadez19, YuchengWang20, Goncalves23, Shimasaki24, Karcher24}.

Meanwhile, measurement---a fundamental concept in quantum mechanics and a pillar of recent developments in quantum technology---also has an effect similar to disorder. 
By repeated applications of measurements in a local basis, particles or spins are frozen into certain configurations randomly selected due to the probabilistic nature of measurement outcomes. 
Analogous to the Anderson transitions, such frequent measurements induce entanglement phase transitions between an area-law and a volume-law entangled phase, which are also accompanied by critical scaling behaviors of various physical quantities \cite{YaodongLi18, Skinner19, AmosChan19, YaodongLi19, Szyniszewski19, Bao20, Jian20, Gullans20} (see also reviews \cite{Potter22, Lunt22, Fisher23}). 
Multifractality of wave functions in the many-body basis has been used to characterize the universal nature of the measurement-induced entanglement transitions in monitored random circuits \cite{Sierant22}. 
It has also been demonstrated that steady-state wave functions, in an appropriate basis, reveal multifractality in the volume-law entangled phases of monitored Clifford circuits and in the critical phase of a nonunitary free-fermion circuit \cite{Iaconis21}. 
While the very existence of entanglement transitions in monitored free-fermion systems, especially with particle-number conservation, remains elusive in one dimension \cite{Cao19, Alberton21, Buchhold21, Fidkowski21, Coppola22, Szyniszewski23, Poboiko23}, an entanglement transition in two dimensions has been found to exhibit multifractality of single-particle wave functions quantitatively similar to an Anderson transition in three dimensions \cite{Chahine23}. 
However, the ubiquity and universal characters of multifractality in monitored quantum systems remain largely unexplored.

In this work, we study multifractality appearing in the probability distribution of a single quantum or classical particle subject to repeated measurements. 
For the quantum case, a single particle is evolved by a quantum circuit consisting of local unitary gates and measurements of the single-site occupation numbers. 
It is worth mentioning that this quantum circuit can be viewed as a $U(1)$-symmetric monitored circuit, which has been studied in the context of measurement-induced phase transition \cite{Agrawal22, Barratt22, Oshima23, Chakraborty23}, with restriction to the Hilbert space of a single particle.
For the classical case, a single particle is evolved by local stochastic processes, and we try to estimate its probability distribution with partial knowledge of the occupation numbers gained by measurements.
In both cases, we examine scaling behaviors for the steady-state values of the averaged inverse participation ratio and find that projective measurements generically induce multifractality. 
However, the nature of multifractality is sharply contrasted with that for Anderson transitions by the lack of scale invariance. 
It also strongly depends on whether the wave-packet dynamics in the absence of measurements is diffusive or ballistic, but is otherwise immune to the strength of measurements [as far as the measurement rate is $p \sim O(1/L)$] or differences in protocol. 
On the other hand, multifractality turns out to be lost by generalized measurements with erroneous outcomes or by postselection of trajectories in which the particle is never detected. 
These results are summarized in Table~\ref{tab:Multifractality}. 
\begin{table*}
\begin{ruledtabular}
\begin{tabular}{lllll}
System & Evolution & Measurement & Multifractality & Section \\ \hline
Quantum & Random unitary gates & Projective & Yes ($\tau_2 \simeq 0.51$) & Sec.~\ref{sec:NumQuantumRandom} \\
& Fixed unitary gates & Projective & Yes ($\tau_2 \simeq 0.79$) & Sec.~\ref{sec:NumQuantumFixed} \\
& Random unitary gates & Generalized & No & Sec.~\ref{sec:NumQuantumGeneral} \\
& Random unitary gates & No-click & No & Sec.~\ref{sec:NumQuantumNoclick} \\ \hline
Classical & Random transition matrices & Projective \footnote{The projective measurements for classical systems mean that measurements do not involve errors.} & Yes ($\tau_2 \simeq 0.56$) & Sec.~\ref{sec:NumClassicalRandom} \\
& Fixed transition matrices & Projective \footnotemark[1] & Yes ($\tau_2 \simeq 0.54$) & Sec.~\ref{sec:NumClassicalFixed}
\end{tabular}
\end{ruledtabular}
\caption{List of circuit models considered in this manuscript. 
When multifractality is present, we also provide the exponent $\tau_2$ for the mean IPR extracted for the measurement rate $p=1/L$.}
\label{tab:Multifractality}
\end{table*}
We further provide simple analytically tractable models---a single-shot measurement model and a random walk subject to stochastic resetting---that capture several essential properties of multifractality caused by measurements.

We note that the dynamics of a single quantum or classical particle subject to repeated measurements has also been studied in Refs.~\cite{Jin22, Jin23, Popperl23}. 
In particular, Ref.~\cite{Popperl23} considered a one-dimensional Anderson chain subject to local projective measurements, for which the particle is exponentially localized in the absence of measurements.
They then investigate delocalizing effects of the measurements for the single-particle dynamics. 
Instead, we here consider localizing effects of projective measurements on a single particle, which tends to spread in the absence of measurements due to the unitary evolution described by local unitary circuits. 
We also intensively focus on quantities nonlinear in the density matrix, such as the inverse participation ratio, whose averages are conditioned on the measurement outcomes, in close analogy with the measurement-induced phase transitions for which the conditional average of nonlinear quantities, such as entanglement entropy, is typically considered. 

The rest of this manuscript is organized as follows. 
In Sec.~\ref{sec:QuantumSystem}, we focus on monitored quantum circuits for a single particle. 
After introducing the model (Sec.~\ref{sec:Model}) and basics on the multifractal scaling analysis (Sec.~\ref{sec:Multifractal}), we provide a toy model of single-shot measurements to explain several basic properties of multifractal exponents (Sec.~\ref{sec:SingleShot}). 
We then present our numerical results for a variety of quantum circuits (Sec.~\ref{sec:NumQuantum}).
In Sec.~\ref{sec:ClassicalSystem}, we numerically investigate multifractal properties of classical counterparts of monitored circuits for a single particle. 
In particular, we point out a similarity between our monitored systems with a diffusive particle and a classical random walk with stochastic resetting in Sec.~\ref{sec:StochasticResetting}.
We conclude our manuscript in Sec.~\ref{sec:Summary}.

\section{Quantum system}
\label{sec:QuantumSystem}

\subsection{Model}
\label{sec:Model}

\begin{figure}
\includegraphics[width=0.45\textwidth]{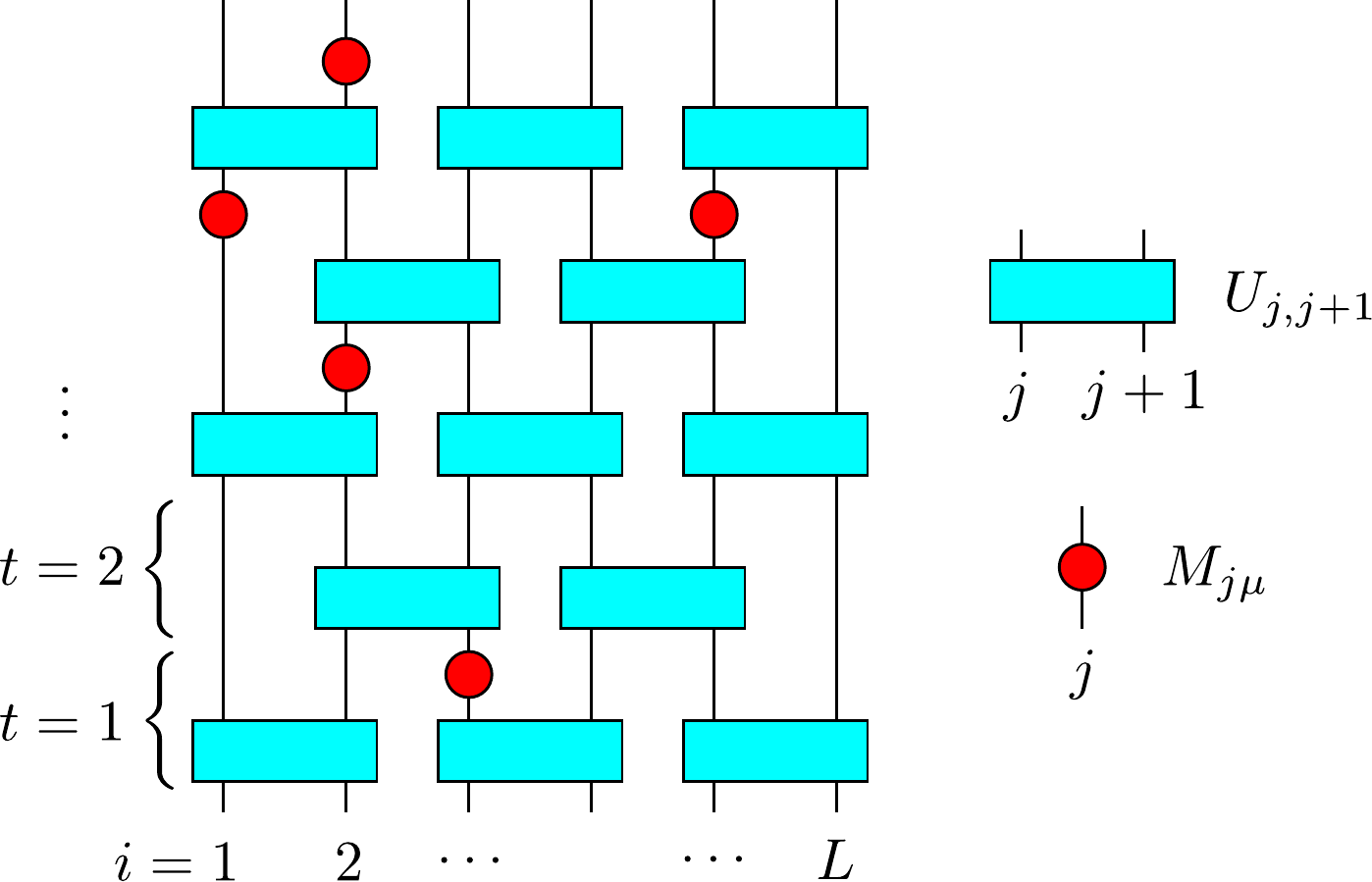}
\caption{Quantum circuit consists of layers of two-site unitary gates $U_{j,j+1}$ (cyan rectangles) and generalized measurements $M_{j\mu}$ (red circles). 
The horizontal and vertical direction corresponds to space and time, respectively.
}
\label{fig:QuantumCircuit}
\end{figure}

We consider a single quantum particle moving on a one-dimensional lattice with the length $L$. 
Let $| i \rangle$ be an orthonormal basis state with a particle localized at site $i \in \{1,2,\cdots,L\}$. 
An arbitrary single-particle quantum state $| \psi \rangle$ is then expanded in this basis as 
\begin{align}
| \psi \rangle = \sum_{i=1}^L c_i | i \rangle
\end{align}
with complex coefficients $c_i$ subject to the normalization condition $\sum_{i=1}^L |c_i|^2 = 1$. 
The quantum state is evolved in time by a quantum circuit as schematically shown in Fig.~\ref{fig:QuantumCircuit}. 
The unit-time evolution for this circuit consists of two steps: (i) the application of a sequence of two-site unitary gates and (ii) measurements of the occupation numbers at every site, as we will explain below in detail. 
In the following discussion, we set the system length $L$ to be even.
We assume that the particle is initially placed at the center of the lattice, i.e., $| \psi(0) \rangle = | L/2 \rangle$. 
We note that the choice of initial state merely changes the transient behavior and does not affect the steady-state properties of the circuit evolution (we have numerically confirmed this for a small-size system).

The first step of the unit-time evolution is a unitary evolution by two-site unitary gates. 
The unitary gates $U_{j,j+1}$ are $2 \times 2$ unitary matrices acting on the basis states $| j \rangle$ and $| j+1 \rangle$. 
A quantum state $| \psi(t) \rangle$ at a time $t$ is evolved by a sequence of random unitary gates as 
\begin{align} \label{eq:UnitaryGates}
| \tilde{\psi}(t) \rangle = \left( \bigoplus_{j \in \mathcal{S}_t} U_{j,j+1} \right) | \psi (t) \rangle,
\end{align}
where $\mathcal{S}_t = \{ 1,3,\cdots,L-1 \}$ for odd $t$, $\mathcal{S}_t = \{ 2,4,\cdots,L-2 \}$ for even $t$ under the open boundary condition (OBC). 
Unless otherwise mentioned, we impose the OBC in the following analysis. 
The periodic boundary condition (PBC), where $\mathcal{S}_t = \{ 2,4,\cdots,L \}$ for even $t$ with a convention $U_{L,L+1} \equiv U_{L,1}$, is used only in cases for investigating averaged properties of the probability distribution itself.
In the following numerical analysis, we basically consider Haar-random unitary matrices for the unitary gates $U_{j,j+1}$, that is, $U_{j,j+1}$ are uniformly drawn from the Haar measure of the unitary group $U(2)$ independently at each time and for each bond $(j,j+1)$. 
The only exception happens in Sec.~\ref{sec:NumQuantumFixed} where we choose $U_{j,j+1}$ to be a single fixed unitary matrix, so that the time evolution becomes deterministic in the absence of measurements.

At the second step of the unit-time evolution, we measure the occupation number at every site with the probability $p \in [0,1]$. 
The measurement at the site $i$ can be a generalized measurement represented by a set of Kraus operators $\{ M_{ik} \}$ where $k$ labels the possible measurement outcome, which is either $0$ or $1$ in this case. 
If the measurement is performed on a state $| \tilde{\psi}(t) \rangle$, the outcome $\mu \in \{ 0,1 \}$ is obtained with the Born probability, 
\begin{align}
P[k = \mu] = \langle \tilde{\psi}(t) | M^\dagger_{i\mu} M_{i\mu} | \tilde{\psi}(t) \rangle. 
\end{align}
The state $| \tilde{\psi}(t) \rangle$ is then updated as 
\begin{align}
| \tilde{\psi}(t) \rangle \to | \tilde{\psi}'(t) \rangle = \frac{M_{i\mu} | \tilde{\psi}(t) \rangle}{\left\| M_{i\mu} | \tilde{\psi}(t) \rangle \right\|}.
\end{align}
Each pure state $| \tilde{\psi}'(t) \rangle$ labeled by a sequence of the measurement outcomes $\{ \mu \}$ is called the quantum trajectory.
In most cases discussed below, we consider projective measurements of the occupation number at each site, which are given by the Kraus operators,
\begin{align} \label{eq:ProjectiveMes}
\begin{split}
M_{i0} &= I -| i \rangle \langle i |, \\
M_{i1} &= | i \rangle \langle i |,
\end{split}
\end{align}
where $I$ is the identity operator.
In Sec.~\ref{sec:NumQuantumGeneral}, we also consider the case for generalized measurements, which are given by
\begin{align} \label{eq:GeneralizedMes}
\begin{split}
M_{i0} &= \sqrt{1-\frac{e}{2}} \left( I -| i \rangle \langle i | \right) +\sqrt{\frac{e}{2}} | i \rangle \langle i |, \\
M_{i1} &= \sqrt{\frac{e}{2}} \left( I -| i \rangle \langle i | \right) +\sqrt{1-\frac{e}{2}} | i \rangle \langle i |.
\end{split}
\end{align}
Here, the real positive parameter $e \in [0,1]$ controls an error in the measurement outcome; 
for $e=0$ the measurement is perfect and reduces to the projective one in Eq.~\eqref{eq:ProjectiveMes}, whereas for $e=1$ it just becomes the identity operation $M_{i\mu} \propto I$.
Once such measurements are performed over all sites with the probability $p$, the unit-time evolution is accomplished. 
We denote the resulting pure state (trajectory) by $| \psi(t+1) \rangle$. 
The above procedure of the unitary and measurement operations is repeated until the time $t=T$ at which quantities of interest after averaging, as we will explain below, reach steady-state values.

In this model, the unitary gates tend to spread an initially localized particle over the lattice, whereas the projective measurements of the occupation number immediately localize the particle once it is detected with the outcome $\mu=1$. 
Our main objective is to show that the competition between the unitary dynamics and measurements can induce nontrivial multifractal scaling properties of the quantum trajectory at the single-particle level.
In fact, when the probability for applying the measurement on each site is of the order of unity, i.e., $p \sim \mathcal{O}(1)$, the projective measurements are too frequent for the particle to spread, rendering the quantum trajectory localized in the large-volume limit $L \to \infty$ for any finite $p$. 
We thus choose the probability $p$ to scale with $1/L$. 
This means that at each time step the measurement is performed $\mathcal{O}(1)$ times on average irrespective of the system size. 
As we will see in Sec.~\ref{sec:NumQuantum}, the projective measurements with $p \sim \mathcal{O}(1/L)$ generally lead to a multifractal behavior in a single-particle quantum trajectory. 
On the other hand, the quantum trajectory only exhibits an extended behavior, as if to evolve under the purely unitary dynamics, for the generalized measurements with erroneous outcomes or after postselecting trajectories in which the particle is never detected (no-click measurements).

\subsection{Multifractal scaling analysis}
\label{sec:Multifractal}

As commonly employed in the multifractal scaling analysis for Anderson transitions \cite{Evers08}, we introduce the inverse participation ratio (IPR) for a single-particle quantum state $| \psi \rangle$ by 
\begin{align}
\textrm{IPR}(q) \equiv \sum_{i=1}^L | \langle i | \psi \rangle |^{2q} = \sum_{i=1}^L |c_i|^{2q},
\end{align}
with $q$ being a real number. 
While the IPR can often be considered for $q \leq 0$ in the literature, the quantum trajectory $| \psi(t) \rangle$ obtained by the above prescription typically has exactly vanishing components $c_i=0$ due to projective measurements, which make the IPR ill-defined for $q \leq 0$. 
We thus restrict ourselves to $q > 0$. 

We focus on the IPR averaged over quantum trajectories $\{ | \psi(t) \rangle \}$. 
In our model of a single quantum particle, it generally involves three types of averaging: (i) the average over random unitary gates, (ii) the average over measurement positions, and (iii) the average over measurement outcomes. 
We denote a quantity $X$ averaged over all these possibilities by $\langle X \rangle$. 
In analogy with Anderson transitions, we expect that the IPR averaged over trajectories (mean IPR) asymptotically scale as 
\begin{align} \label{eq:AverageIPR}
\langle \textrm{IPR}(q) \rangle \sim L^{-\tau_q}.
\end{align}
The exponent $\tau_q$ is thus defined by 
\begin{align} \label{eq:ExponentTau}
\tau_q = -\lim_{L \to \infty} \frac{\ln \langle \textrm{IPR}(q) \rangle}{\ln L}.
\end{align}
The fractal dimension $D_q$ defined through $\tau_q \equiv D_q (q-1)$ quantifies the multifractal behavior of the ensemble of quantum trajectories. 
The trajectories are diagnosed to be localized when $D_q = 0$ whereas extended when $D_q = 1$. 
When the trajectories are multifractal, we expect that $D_q$ becomes some nontrivial function of $q$.

We also introduce another exponent $\tau_q^*$, which is related to the typical IPR,
\begin{align} \label{eq:TypicalIPR}
e^{\langle \ln \textrm{IPR}(q) \rangle} \sim L^{-\tau^*_q}.
\end{align} 
It is defined by changing the order of average and logarithm in Eq.~\eqref{eq:ExponentTau}, 
\begin{align}
\tau_q^* = -\lim_{L \to \infty} \frac{\langle \ln \textrm{IPR}(q) \rangle}{\ln L}.
\end{align}
In the context of Anderson transitions, the exponents $\tau_q$ and $\tau_q^*$ are expected to coincide with each other when the system is off critical (either extended or localized), while they can differ from each other when the system is critical with multifractality \cite{Evers00, Mirlin00}. 
As we will see below, projective measurements make the behaviors of these two exponents different for $q \gtrsim 2$. 

In addition to the IPR, we focus on the variance of the position operator $x \equiv \sum_{i=1}^L i | i \rangle \langle i |$, 
\begin{align}
\textrm{Var} = \langle \psi(t) | x^2 | \psi(t) \rangle -\langle \psi(t) | x | \psi(t) \rangle^2,
\end{align}
whose trajectory average is also expected to asymptotically scale as $\langle \textrm{Var} \rangle \sim L^{2\tau_\textrm{Var}}$. 
We expect that $\tau_\textrm{Var}=1$ when the trajectories are extended whereas $\tau_\textrm{Var}=0$ when they are localized. 

We remark that the density matrix $\rho(t) = | \psi(t) \rangle \langle \psi(t) |$ averaged over quantum trajectories reaches the infinite-temperature state $\langle \rho(t) \rangle \propto I$ after a sufficiently long time, regardless of the measurement rate $p$. 
If the probability distribution $| c_i(t) |^2 = \textrm{Tr}[|i \rangle \langle i | \rho(t)]$ is averaged over trajectories, it simply becomes a uniform distribution $\langle |c_i(t)|^2 \rangle = 1/L$. 
Thus, a nontrivial multifractal behavior is not revealed in the IPR computed from the moments of such an averaged probability distribution $\langle |c_i(t)|^2 \rangle$.

\subsection{Toy model of single-shot measurements}
\label{sec:SingleShot}

Before going to details of our numerical results, we consider a simple tractable model of monitored single-particle dynamics.
We suppose the uniform probability distribution for the particle density,
\begin{align}
p_i = |c_i|^2 = \frac{1}{L},
\end{align}
which could be approximately realized by applying a unitary circuit with a sufficiently long depth on an arbitrary single-particle state $| \psi \rangle$. 
We then perform the projective measurement $\{ M_{jk} \}$ of the occupation number at a randomly chosen site $j$. 
The particle will be detected with the probability $P[k=1]=1/L$ and the resulting probability distribution collapses into a perfectly localized one, $p_i = \delta_{ij}$, leading to $\textrm{IPR}(q)=1$. 
On the other hand, the particle will be undetected with the probability $P[k=0]=1-1/L$. 
In this case, we find the probability distribution $p_i=(1-\delta_{ij})/(L-1)$ and $\textrm{IPR}(q) = (L-1)^{1-q}$. 
We then consider the IPR averaged over trajectories generated by this single-shot projective measurement on the uniformly distributed single-particle state.

For the mean IPR in Eq.~\eqref{eq:AverageIPR}, we find
\begin{align}
\langle \textrm{IPR}(q) \rangle = \frac{1+(L-1)^{2-q}}{L}
\end{align}
and thus
\begin{align} \label{eq:TauSingleShot}
\tau_q = \begin{cases} q-1 & (q < 2) \\ 1 & (q \geq 2) \end{cases}.
\end{align}
On the other hand, for the typical IPR in Eq.~\eqref{eq:TypicalIPR}, we have 
\begin{align}
\langle \ln \textrm{IPR}(q) \rangle = \frac{L-1}{L} (1-q) \ln (L-1)
\end{align}
and thus
\begin{align} \label{eq:Tau*SingleShot}
\tau_q^* = q-1.
\end{align}
This indicates that the exponent $\tau_q$ for the mean IPR is more sensitive to the presence of rare trajectories with a localized particle, which dominates the large-$q$ behavior. 

This analysis can be readily extended to the projective measurements of the occupation numbers at multiple sites. 
If we measure $r$ sites on the uniformly distributed single-particle state, we find for the mean IPR, 
\begin{align}
\langle \textrm{IPR}(q) \rangle = \frac{r +(L-r)^{2-q}}{L},
\end{align}
and for the typical IPR, 
\begin{align}
\langle \ln \textrm{IPR}(q) \rangle = \frac{L-r}{L} (1-q) \ln (L-r).
\end{align}
Therefore, the above results for $\tau_q$ and $\tau_q^*$ are not altered by the projective measurements of multiple sites as long as $r \sim \mathcal{O}(1)$. 
As we will see in the next section, the repeated applications of unitary gates and projective measurements with a probability $p \sim \mathcal{O}(1/L)$ render both $\tau_q$ and $\tau_q^*$ smooth, nontrivial functions of $q$. 
While they almost coincide with each other for small $q$, a strong deviation occurs for $q \gtrsim 2$ and, in particular, $\tau_q$ saturates to $\tau_\infty = 1$ for large $q$ as anticipated from the single-shot measurement result. 

When the projective measurements are performed at each site with a probability $p \sim \mathcal{O}(1)$, we will have $r \sim \mathcal{O}(L)$. 
In this case, we can still obtain nontrivial exponents corresponding to delocalized trajectories. 
However, we expect that many rounds of local unitary gates and projective measurements eventually make the trajectories localized and only give the trivial exponents $\tau_q = \tau_q^* = 0$ in the steady-state regime.

Although the projective measurements lead to the distinct behaviors between the mean and typical IPRs, the generalized measurements with possible erroneous outcomes have dramatic effects on the mean IPR. 
When the particle is ``detected'' by the generalized measurements defined in Eq.~\eqref{eq:GeneralizedMes}, the IPR for the corresponding state becomes 
\begin{align}
\textrm{IPR}(q) = \frac{(2-e)^q +(L-1) e^q}{[(L-1)e+2-e]^q},
\end{align}
which asymptotically scales as $\textrm{IPR}(q) \sim L^{1-q}$ for any finite error rate $e$. 
This is the same asymptotic form as those for the ``undetected'' case and also for extended trajectories. 
Therefore, both mean and typical IPRs have the same scaling behavior in the large-volume limit with $\tau_q = \tau_q^* = q-1$. 
This indicates that any finite error rate renders the trajectories extended after a single shot of measurements. 
This tendency can also be seen in the quantum circuit consisting of local unitary gates and generalized measurements as we will discuss in Sec.~\ref{sec:NumQuantumGeneral}.

\subsection{Numerical results}
\label{sec:NumQuantum}

We here numerically investigate the multifractal scaling properties of single-particle quantum trajectories evolved under the monitored circuits presented in Sec.~\ref{sec:Model}. 
We consider several combinations of local unitary gates and measurement schemes: Haar random unitary gates and projective measurements (Sec.~\ref{sec:NumQuantumRandom}), fixed unitary gates and projective measurements (Sec.~\ref{sec:NumQuantumFixed}), Haar random unitary gates and generalized measurements with possible erroneous outcomes (Sec.~\ref{sec:NumQuantumGeneral}), and Haar random unitary gates and no-click measurements by postselecting trajectories in which the particle is never detected (Sec.~\ref{sec:NumQuantumNoclick}). 
We find nontrivial multifractal scaling of the IPR for the first two cases with projective measurements, whereas the scaling becomes simply extended for the last two cases with the generalized or no-click measurements.

\subsubsection{Random unitary + projective measurements}
\label{sec:NumQuantumRandom}

\begin{figure}
\includegraphics[width=0.47\textwidth]{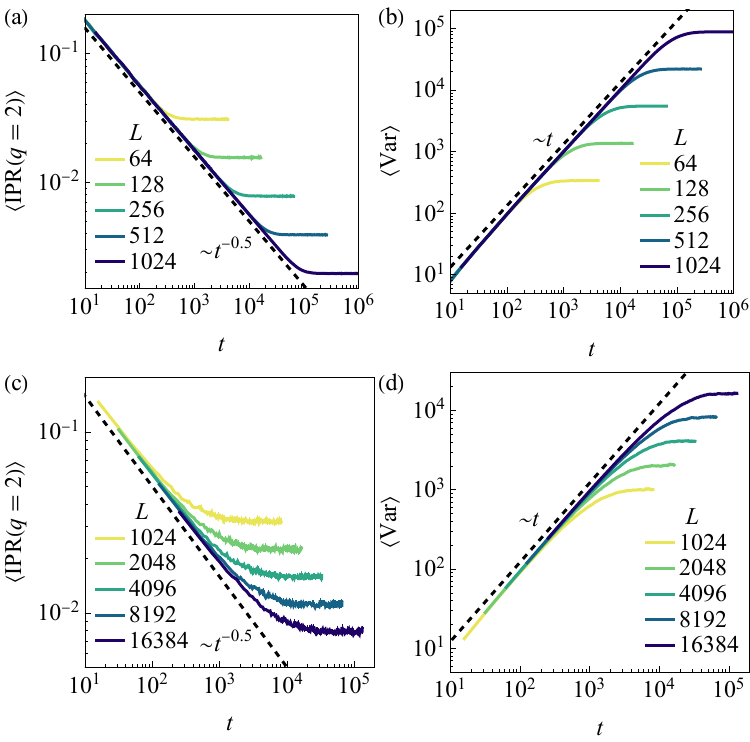}
\caption{Time evolution of the mean IPR for $q=2$ and the variance of the position operator in the quantum circuit composed of Haar random unitary gates and projective measurements. 
The results are shown for the measurement rate $p=0$ in (a) and (b) and for $p=1/L$ in (c) and (d). 
}
\label{fig:TimeEvolQuantumRandom}
\end{figure}

We first consider the quantum circuit consisting of Haar random unitary gates and projective measurements of the occupation numbers. 
As discussed in Sec.~\ref{sec:Model}, the projective measurements are applied at every site with the probability $p \sim \mathcal{O}(1/L)$ so that we only measure $\mathcal{O}(1)$ sites at each time step. 
In Figs.~\ref{fig:TimeEvolQuantumRandom}(a) and (b), we show time evolution of the mean IPR for $q=2$ and the variance of the position operator, respectively, for $p=0$, both of which are averaged over 1000 trajectories. 
The IPR decays as $t^{-1/2}$ whereas the variance grows as $t$, clearly indicating diffusive spreading of the wave packet in the random circuit evolution. 
In Figs.~\ref{fig:TimeEvolQuantumRandom}(c) and (d), we show time evolution of the same quantities in the presence of projective measurements with $p=1/L$, averaged over 8000 trajectories, which initially decay or grow in the same rates as in the $p=0$ case but reach steady-state values at much faster times. 
We hereafter focus on the steady-state values of the averaged IPR or variance computed at $t=L^2$ for $p=0$ and $t=8L$ for $p > 0$.

\begin{figure}
\includegraphics[width=0.47\textwidth]{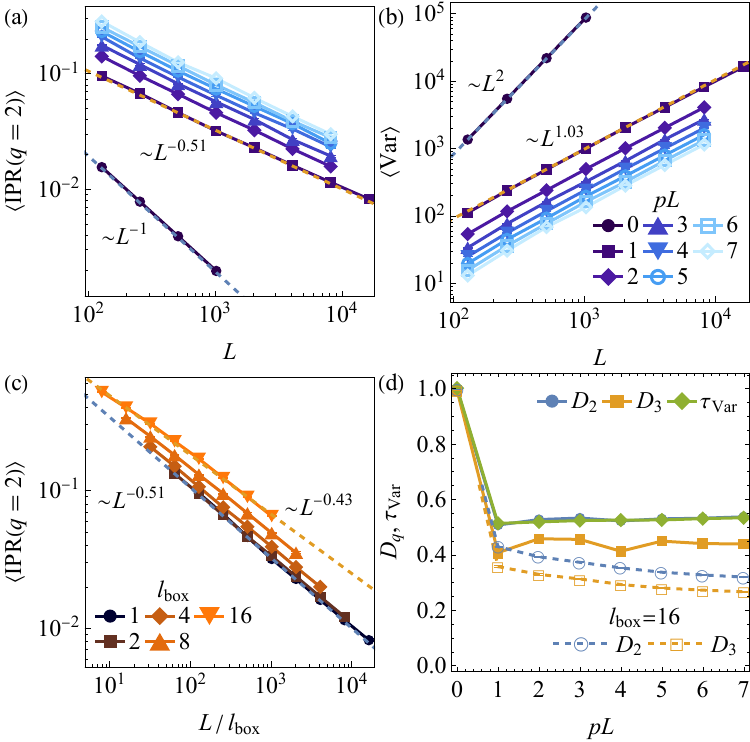}
\caption{Steady-state values of (a) the mean IPR for $q=2$ and (b) the variance of the position operator as functions of $L$ in the quantum circuit composed of Haar random unitary gates and projective measurements. 
(c) Mean IPR for coarse-grained probability distributions with the box size $l_\textrm{box}$.
The dashed lines are fitting functions.
(d) The fractal dimension $D_q$ for $q=2,3$ and exponent $\tau_\textrm{Var}$ are extracted from data with $L \geq 128$ for $l_\textrm{box}=1$ (filled symbols) and $l_\textrm{box}=16$ (empty symbols) and plotted against the measurement rate $p$. 
}
\label{fig:IPRQuantumRandom}
\end{figure}

In Figs.~\ref{fig:IPRQuantumRandom}(a) and (b), we show the steady-state values of the mean IPR for $q=2$ and variance of the position operator, which are plotted against the system size $L$. 
Here, the averages are taken over 1000 trajectories for $p=0$ and 8000 trajectories for $p>0$. 
Both IPR and variance show power-law scaling forms, but their exponents take nontrivial values $\tau_2 \simeq \tau_\textrm{Var} \simeq 0.5$ in the presence of projective measurements, which are clearly distinct from the values $\tau_2 = \tau_\textrm{Var} = 1$ expected for extended trajectories in the absence of measurements. 
This implies that repeated projective measurements with the rate $p \sim \mathcal{O}(1/L)$ hinder spreading of the particle but are insufficient to localize the particle, thereby signaling multifractality in the ensemble of quantum trajectories. 
However, in contrast to critical systems at Anderson transitions \cite{Janssen94, Vasquez08, Rodriguez08}, scaling properties of the IPR are affected by coarse graining. 
In order to see this, we partition the system into $L/l_\textrm{box}$ boxes of the size $l_\textrm{box}$ and consider a coarse-grained probability distribution, 
\begin{align}
\mu_k (l_\textrm{box}) = \sum_{i=1}^{l_\textrm{box}} \left| \langle i+kl_\textrm{box} | \psi \rangle \right|^2.
\end{align}
We then compute the IPR with respect to this distribution by 
\begin{align}
\textrm{IPR}(q) = \sum_{k=0}^{L/l_\textrm{box}-1} \mu^q_k(l_\textrm{box}).
\end{align}
As shown in Fig.~\ref{fig:IPRQuantumRandom}(c), the mean IPRs computed for various box sizes $l_\textrm{box}$ do not collapse into a single curve, and have the exponent $\tau_2$ varying with $l_\textrm{box}$ from $0.51$ ($l_\textrm{box}=1$) to $0.43$ ($l_\textrm{box}=16$). 
In Fig.~\ref{fig:IPRQuantumRandom}(d), we show the fractal dimension $D_q = \tau_q / (q-1)$ for $q=2,3$ and the exponent $\tau_\textrm{Var}$ as functions of $p$, which are obtained by least-squares fitting for data with $L \geq 128$.
The exponents generically take nontrivial values between 0 and 1 in the presence of projective measurements. 
The discrepancy between $D_2$ and $D_3$ strongly indicates the nonlinearity of $\tau_q$ and thus the emergence of multifractality. 
Furthermore, the exponents do not much depend on the measurement rate $p$, except the slow decreases of $D_q$ for $l_\textrm{box}=16$. 
In short, we found that the exponent $\tau_q$ for the mean IPR is fragile to coarse graining, but it is rather insensitive to the measurement rate $p$ once the box size $l_\textrm{box}$ is fixed as long as $p\sim\mathcal{O}(1/L)$. 

\begin{figure}
\includegraphics[width=0.47\textwidth]{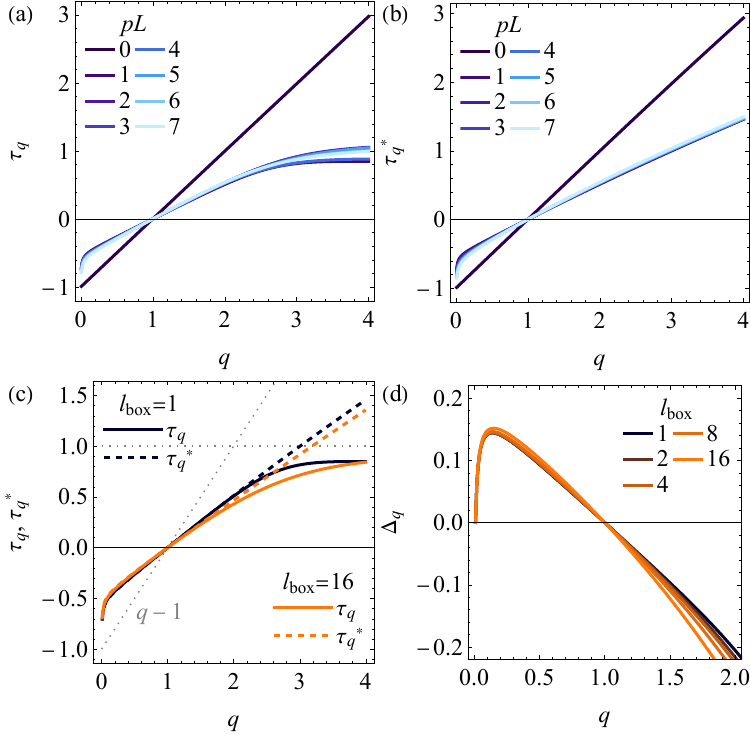}
\caption{Exponents for (a) the mean IPR and (b) the typical IPR as functions of $q$ in the quantum circuit composed of Haar random unitary gates and projective measurements. 
(c) The two exponents for $p=1/L$ are compared between the original one (black lines), coarse-grained one (orange lines), and single-shot measurement results (dotted lines). 
(d) Anomalous dimension $\Delta_q$ defined through $\tau_q = D_0(q-1)+\Delta_q$.
}
\label{fig:ExponentQuantumRandom}
\end{figure}

By performing the fitting analysis for various values of $q \leq 4$, we obtain the exponent $\tau_q$ for the mean IPR and $\tau_q^*$ for the typical IPR as functions of $q$. 
The results are shown in Figs.~\ref{fig:ExponentQuantumRandom}(a) and (b).
In the absence of measurements, both exponents take functional forms very close to $\tau_q = q-1$ as expected for extended trajectories. 
In the presence of projective measurements, the exponents become smooth, nonlinear functions of $q$ and only show weak dependence on the measurement rate $p$ within the regime $pL \leq 7$ we are focusing on. 
In Fig.~\ref{fig:ExponentQuantumRandom}(c), we compare the exponents $\tau_q$ and $\tau_q^*$ for $p=1/L$ and those obtained in the single-shot measurement model in Sec.~\ref{sec:SingleShot}. 
While the functional forms of the exponents are different between the circuit model and the single-shot measurement model, both models have common features that the two exponents almost coincide with each other for $q \lesssim 2$ and $\tau_q$ saturates to $\tau_\infty = 1$ for sufficiently large $q$. 
This indicates that multifractality of the trajectories is revealed in both $\tau_q$ and $\tau_q^*$, but the exponent $\tau_q$ for the mean IPR is more sensitive to rare trajectories with a localized particle. 

\begin{figure}
\includegraphics[width=0.47\textwidth]{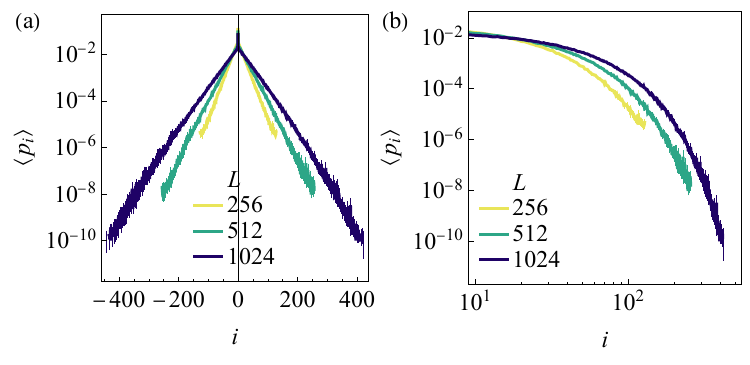}
\caption{Averaged probability distributions for the quantum circuit composed of Haar random unitary gates and projective measurements under the PBC with adjusted localization centers are shown in (a) log-linear and (b) log-log scale. 
The averages are taken over 4000 trajectories for $p=1/L$.}
\label{fig:PDQuantumRandom}
\end{figure}

We then extract the anomalous dimension $\Delta_q$ defined through $\tau_q = D_0 (q-1) +\Delta_q$, which is shown in Fig.~\ref{fig:ExponentQuantumRandom}(d) for various box sizes $l_\textrm{box}$. 
Here, the fractal dimension $D_0$ is estimated via $D_0 = \tau_q/(q-1)$ at $q=0.01$ (the smallest value of $q$ considered in our numerical analysis) and is found to take values $D_0 \simeq 0.7$ in the presence of measurements, irrespective of the box size. 
The deviation from $D_0=1$ in the standard one-dimensional disordered systems might be attributed to exact zeros in the probability distributions produced by projective measurements.
In contrast to critical disordered systems \cite{Mirlin06, Gruzberg11, Gruzberg13} or several weakly monitored many-body systems \cite{Iaconis21}, the anomalous dimension is unlikely to have the symmetric form $\Delta_q = \Delta_{1-q}$. 
Such multifractality with asymmetric $\Delta_q$ has also been found in some periodically driven systems or disordered systems on random graphs \cite{Bilen21}. 
One origin of the asymmetry is Gaussian fluctuations at small scales, which can be eliminated by coarse graining. 
However, as shown in Fig.~\ref{fig:ExponentQuantumRandom}(d), increasing the box size $l_\textrm{box}$ does not alter the asymmetric shape of $\Delta_q$. 
Another origin is the power-law localization, which has in fact been found in a one-dimensional disordered free fermion system subject to continuous monitoring of the occupation numbers \cite{Szyniszewski23}. 
To test this possibility, we adopt the PBC and consider an averaged probability distribution obtained by adjusting the localization center of each trajectory, determined by the site $i \in \{ -L/2+1, \cdots,L/2 \}$ maximizing $p_i$, to be $i=0$. 
The results are shown in Fig.~\ref{fig:PDQuantumRandom}. 
The averaged probability distribution clearly indicates an exponential localization. 
Hence, neither Gaussian fluctuations nor power-law localization can be the origin of the asymmetric $\Delta_q$. 
Meanwhile, this exponential distribution bears a close resemblance with the stationary distribution for a classical random walk with Poissonian resetting \cite{Evans11, Evans20}. 
As discussed in Sec.~\ref{sec:StochasticResetting}, the random walk indeed well describes the qualitative behavior of the IPR in monitored circuits for a single diffusively spreading particle, including the classical monitored circuits discussed in Sec.~\ref{sec:NumClassical}.

\subsubsection{Fixed unitary + projective measurements}
\label{sec:NumQuantumFixed}

\begin{figure}
\includegraphics[width=0.47\textwidth]{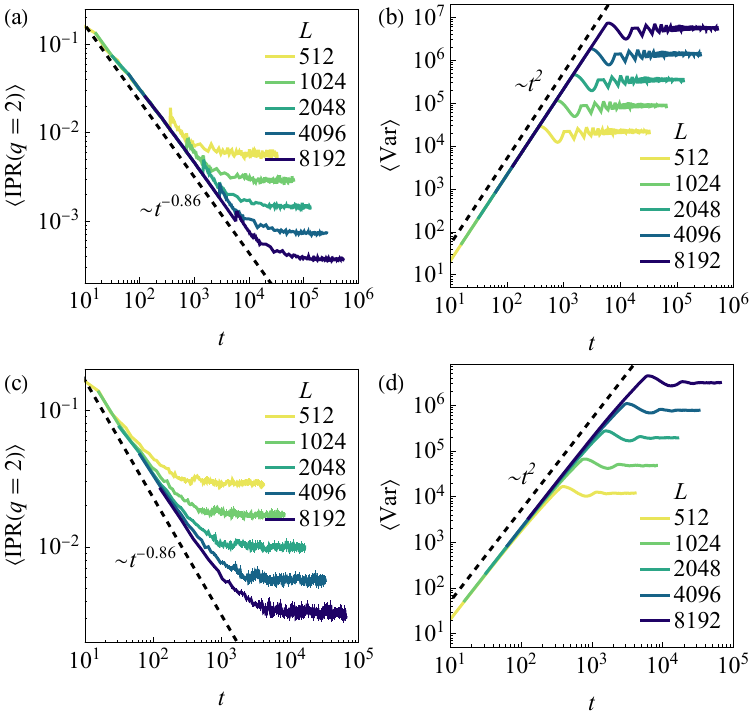}
\caption{Time evolution of the mean IPR for $q=2$ and the variance of the position operator in the quantum circuit composed of fixed unitary gates and projective measurements. 
The results are shown for the measurement rate $p=0$ in (a) and (b) and for $p=1/L$ in (c) and (d).
}
\label{fig:TimeEvolQuantumFixed}
\end{figure}

We here consider a single particle evolved under the quantum circuit consisting of fixed unitary gates and projective measurements.
Specifically, we chose the unitary gate to be
\begin{align} \label{eq:Hadamard}
U_{j,j+1} = \frac{1}{\sqrt{2}} \begin{pmatrix} 1 & 1 \\ -1 & 1 \end{pmatrix}.
\end{align}
Thus, the dynamics becomes completely deterministic in the absence of the measurements. 
In the presence of the projective measurements, the averages of physical quantities are taken over (ii) measurement positions and (iii) measurement outcomes. 
In Figs.~\ref{fig:TimeEvolQuantumFixed}(a) and (b), we show time evolution of the mean IPR for $q=2$ and the variance of the position operator, respectively, for $p=0$.
The growth of the variance with $t^2$ indicates ballistic spreading of the particle in the fixed unitary circuit, in contrast to diffusive spreading found in the random unitary circuit. 
In Figs.~\ref{fig:TimeEvolQuantumFixed}(c) and (d), we show time evolution of the IPR and variance, both of which are averaged over 8000 trajectories, in the presence of projective measurements with $p=1/L$. 
While the early-time scaling behaviors are not much affected by the measurements, both quantities reach the steady-state regime faster in the presence of measurements. 
In the following analysis, we focus on the steady-state values of the IPR or variance computed at $t=64L$ for $p=0$ and at $t=8L$ for $p>0$.

\begin{figure}
\includegraphics[width=0.47\textwidth]{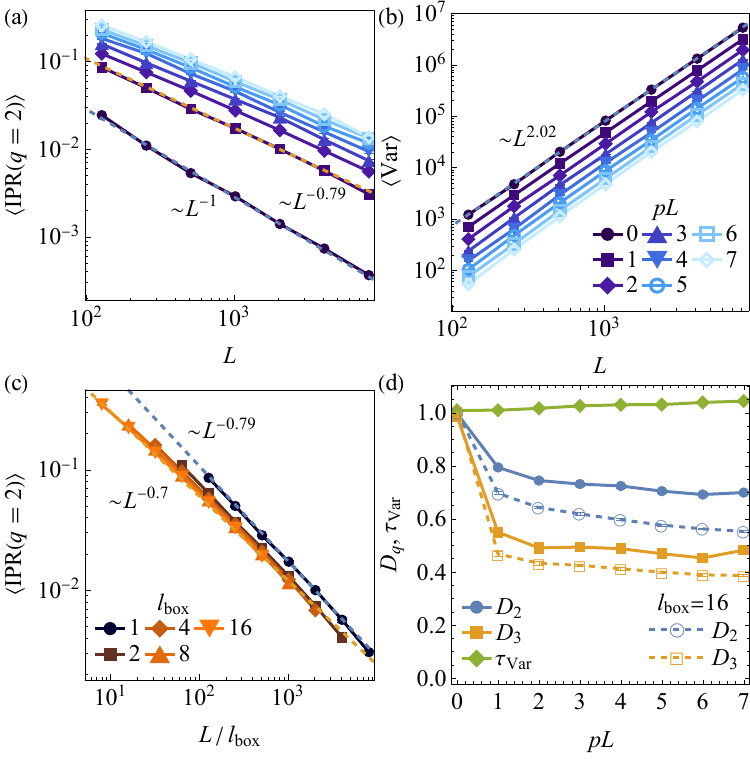}
\caption{Steady-state values of (a) the mean IPR for $q=2$ and (b) the variance of the position operator as functions of $L$ in the quantum circuit composed of fixed unitary gates and projective measurements.
(c) Mean IPR for coarse-grained probability distributions with the box size $l_\textrm{box}$.
The dashed lines are fitting functions.
(d) The fractal dimension $D_q$ for $q=2,3$ and exponent $\tau_\textrm{Var}$ are extracted from the data with $L \geq 128$ for $l_\textrm{box}=1$ (filled symbols) and $l_\textrm{box}=16$ (empty symbols) and plotted against the measurement rate $p$. 
}
\label{fig:IPRQuantumFixed}
\end{figure}

In Figs.~\ref{fig:IPRQuantumFixed}(a) and (b), we show the steady-state values of the mean IPR for $q=2$ and the variance of the position operator, respectively, as functions of the system size $L$.
Here, the averages are taken over 8000 trajectories for $p>0$. 
They show power-law scaling forms with $\tau_2 \simeq \tau_\textrm{Var} \simeq 1$ in the absence of measurements as expected for extended trajectories. 
While the exponent $\tau_\textrm{Var}$ for the variance is almost unaffected by the projective measurements, the exponent $\tau_2$ for the mean IPR is altered and takes a distinct value from the random unitary case. 
The nontrivial value of $\tau_2$ signals multifractality in the ensemble of single-particle trajectories, but, as shown in Fig.~\ref{fig:IPRQuantumFixed}(c), coarse graining of the probability distribution into boxes of the size $l_\textrm{box}$ gradually changes the exponent $\tau_2$ as found in the random unitary case.
In Fig.~\ref{fig:IPRQuantumFixed}(d), we show the fractal dimension $D_q = \tau_q/(q-1)$ for $q=2,3$ and the exponent $\tau_\textrm{Var}$, which are extracted by least-squares fitting for data with $L \geq 128$, as functions of the measurement rate $p$. 
While the fractal dimensions for the coarse-grained distribution with $l_\textrm{box}=16$ slowly decrease with $p$, the exponents do not much depend on the measurement rate.

\begin{figure}
\includegraphics[width=0.47\textwidth]{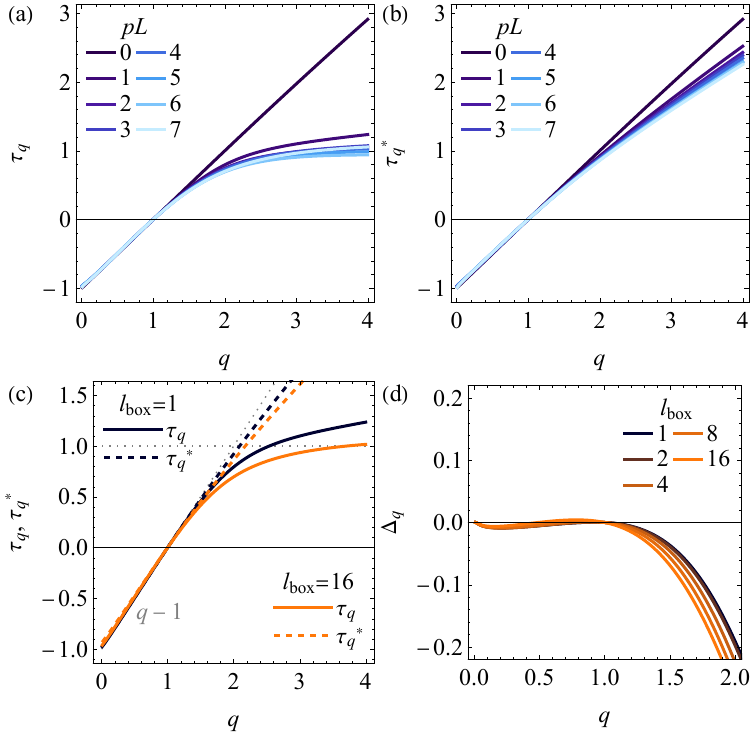}
\caption{Exponents for (a) the mean IPR and (b) the typical IPR as functions of $q$ in the quantum circuit composed of fixed unitary gates and projective measurements. 
(c) The two exponents for $p=1/L$ are compared between the original one (black lines), coarse-grained one (orange lines), and single-shot measurement results (dotted lines). 
(d) Anomalous dimension $\Delta_q$ defined through $\tau_q = D_0(q-1)+\Delta_q$.
}
\label{fig:ExponentQuantumFixed}
\end{figure}

\begin{figure}
\includegraphics[width=0.47\textwidth]{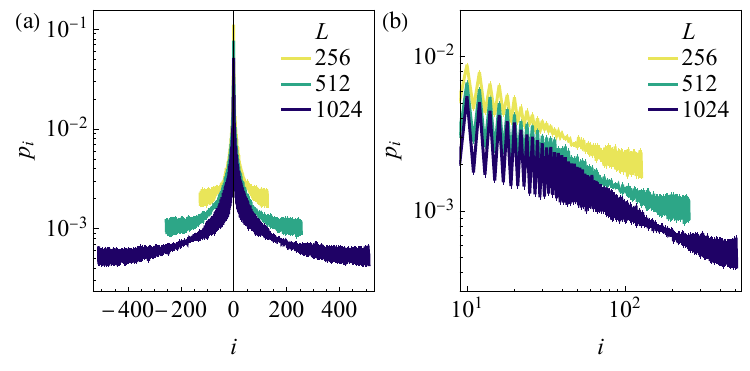}
\caption{Averaged probability distributions for the quantum circuit composed of fixed unitary gates and projective measurements under the PBC with adjusted localization centers are shown in (a) log-linear and (b) log-log scale. 
The averages are taken over 4000 trajectories for $p=1/L$.}
\label{fig:PDQuantumFixed}
\end{figure}

To further confirm multifractality of the quantum trajectories, we obtain the exponent $\tau_q$ for the mean IPR and $\tau_q^*$ for the typical IPR as functions of $q$ by performing the fitting analysis for $q \leq 4$. 
As shown in Fig.~\ref{fig:ExponentQuantumFixed}(a), $\tau_q$ strongly deviates from the linear functional form $q-1$ in the presence of measurements for $q \gtrsim 2$, although the nonlinearity seen around $q=0$ in the random unitary case clearly disappears in the present case. 
On the other hand, as shown in Fig.~\ref{fig:ExponentQuantumFixed}(b), $\tau_q^*$ for the typical IPR only shows small deviation from $q-1$ for $q \gtrsim 2$. 
These features are not affected by coarse graining as compared between the $l_\textrm{box}=1$ and $l_\textrm{box}=16$ cases in Fig.~\ref{fig:ExponentQuantumFixed}(c), while the saturating behavior to $\tau_\infty = 1$, expected from the single-shot measurement model, is more pronounced for $l_\textrm{box}=16$. 
As shown in Fig.~\ref{fig:ExponentQuantumFixed}(d), the absence of nonlinearity around $q=0$ gives $D_0 \simeq 1$ and makes the anomalous dimension $\Delta_q$ flat for $q \lesssim 1$ in contrast to the random unitary case. 
This small-$q$ behavior is due to broad tails of the probability distribution caused by ballistic spreading of the particle, which cannot be seen in slower dynamics by diffusive spreading of the particle in the random unitary case. 
Indeed, algebraic tails can clearly be seen in an averaged probability distribution with adjustment of localization centers to $i=0$ as presented in Fig.~\ref{fig:PDQuantumFixed}.
The discrepancy between $\tau_q$ and $\tau_q^*$ for large $q$ is due to rare localized trajectories caused by projective measurements and has the same origin as in the random unitary case.

\subsubsection{Random unitary + generalized measurements}
\label{sec:NumQuantumGeneral}

\begin{figure}
\includegraphics[width=0.47\textwidth]{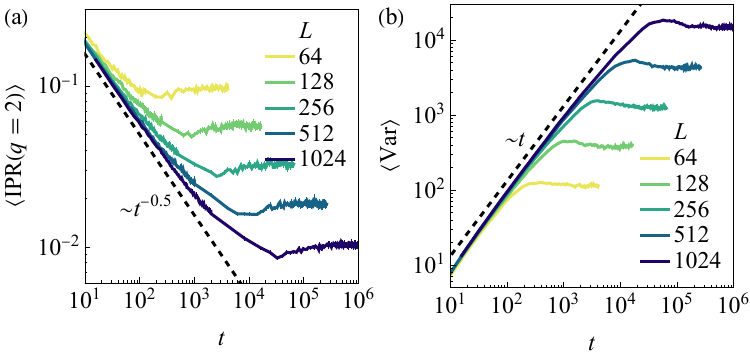}
\caption{Time evolution of (a) the mean IPR for $q=2$ and (b) the variance of the position operator in the quantum circuit composed of Haar random unitary gates and generalized measurements with $p=1/L$ and $e=0.1$. 
}
\label{fig:TimeEvolQuantumGeneral}
\end{figure}

We here consider the quantum circuit consisting of Haar random unitary gates and generalized measurements with possible erroneous measurement outcomes, which turn out to destabilize multifractality found in the circuits with projective measurements.
The generalized measurements are described by the Kraus operators in Eq.~\eqref{eq:GeneralizedMes} and are still applied at every site with the probability $p \sim \mathcal{O}(1/L)$.
The errors in the measurement outcomes are controlled by a real parameter $e \in [0,1]$ and the projective measurements considered in Sec.~\ref{sec:NumQuantumRandom} are recovered when $e=0$. 
As in the case of projective measurements, we consider steady-state quantities averaged over (i) Haar random unitary gates, (ii) measurement positions, and (iii) measurement outcomes. 
In Figs.~\ref{fig:TimeEvolQuantumGeneral}(a) and (b), we show time evolution of the mean IPR and the variance of the position operator, respectively, for $p=1/L$ and $e=0.1$.
Here, the averages are taken over 1000 trajectories. 
As in the case of projective measurements, both quantities reach a steady-state regime after initial diffusive decay or growth. 
In the following analysis, we fix the measurement rate to be $p=1/L$ and focus on the steady-state values of the IPR or variance computed at $t=L^2$. 

\begin{figure}
\includegraphics[width=0.47\textwidth]{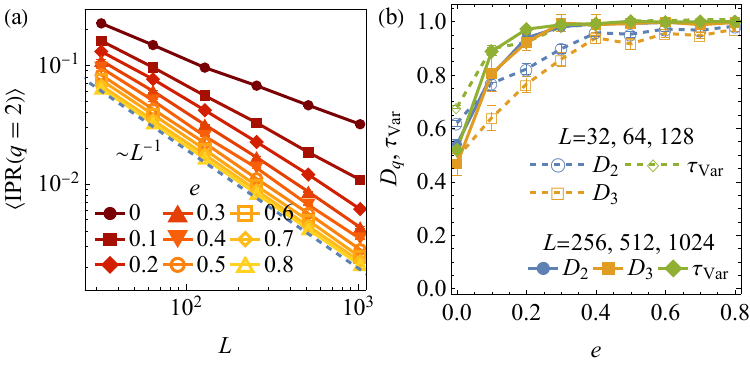}
\caption{(a) Steady-state values of the mean IPR for $q=2$ as functions of $L$ in the quantum circuit composed of Haar random unitary gates and generalized measurements with $p=1/L$.
(b) Fractal dimension $D_q$ for $q=2,3$ and exponent $\tau_\textrm{Var}$ for $p=1/L$ as functions of the error rate $e$. 
They are extracted either from the data set $L=32,64,128$ (dashed lines with empty symbols) or that with $L=256,512,1024$ (solid lines with filled symbols). 
}
\label{fig:IPRQuantumGeneral}
\end{figure}

In Fig.~\ref{fig:IPRQuantumGeneral}(a), we show the steady-state values of the mean IPR for $q=2$ and for various error rates as functions of the system length $L$.
As the dynamics becomes almost unitary for a large error rate, the mean IPR approaches $L^{-1}$ as $e$ is increased. 
As suggested by the analysis on the single-shot measurement model in Sec.~\ref{sec:SingleShot}, any finite error rate $e$ could render trajectories extended and the exponent $\tau_q$ for the IPR may just reduce to the linear function $\tau_q = q-1$ in the large-volume limit. 
In order to see this tendency, we split the data points into two groups, one for small-size systems with $L=32,64,128$ and the other with large-size systems with $L=256,512,1024$, and then perform the least-squares fitting for each group to extract the fractal dimension $D_q$ for the mean IPR and the exponent $\tau_\textrm{Var}$ for the variance of the position operator.
As shown in Fig.~\ref{fig:IPRQuantumGeneral}(b), all these quantities get closer to 1 for the data set including larger system sizes.
We thus expect that any finite error rate destabilizes multifractality of the quantum trajectories observed in the case of projective measurements. 

\subsubsection{Random unitary + no-click measurements}
\label{sec:NumQuantumNoclick}

\begin{figure}
\includegraphics[width=0.47\textwidth]{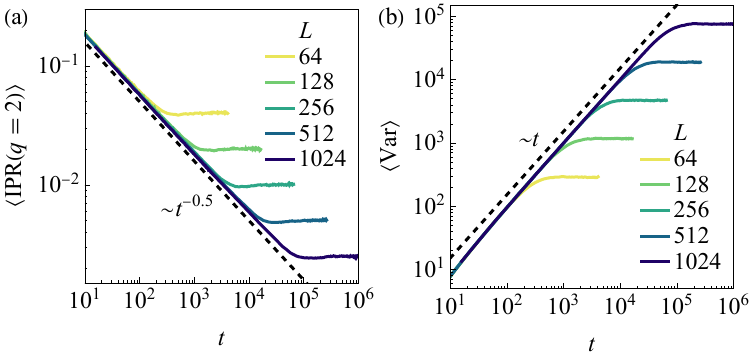}
\caption{Time evolution of (a) the mean IPR for $q=2$ and (b) the variance of the position operator in the quantum circuit composed of Haar random unitary gates and no-click measurements with $p=1/L$.
}
\label{fig:TimeEvolQuantumNoclick}
\end{figure}

As a final example of the quantum circuits for a single particle, we again consider a circuit consisting of Haar random unitary gates and projective measurements, but here we postselect trajectories in which the particle is never detected (i.e., the only $M_{i0}$ is applied). 
We remark that even in this no-click measurement protocol, frequent measurements of the occupation numbers with the probability $p \sim \mathcal{O}(1)$ at every site can eventually locate the position of the particle and render the trajectories localized. 
Thus, we consider the no-click projective measurements applied with the probability $p \sim \mathcal{O}(1/L)$ as in the other cases.
In this setup, the steady-state quantities are averaged over (i) Haar random unitary gates and (ii) measurement positions. 
In Figs.~\ref{fig:TimeEvolQuantumNoclick}(a) and (b), we show time evolution of the mean IPR for $q=2$ and the variance of the position operator, respectively, for $p=1/L$. 
The averages are taken over 1000 trajectories.
They clearly show diffusive spreading of the particle as if the measurements were absent. 
In the following analysis, we focus on steady-state quantities computed at $t=L^2$.

\begin{figure}
\includegraphics[width=0.47\textwidth]{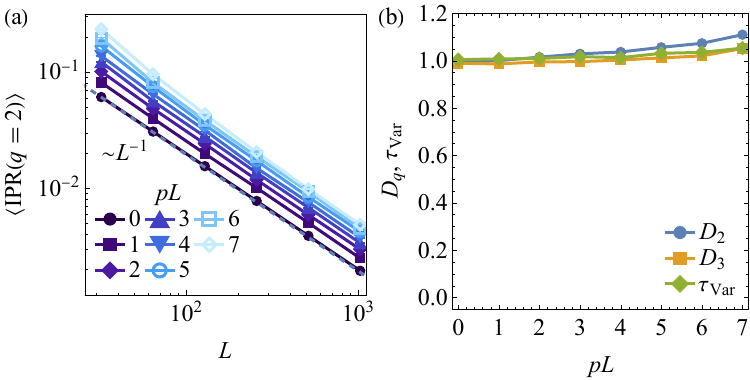}
\caption{(a) Steady-state values of the mean IPR for $q=2$ as functions of $L$ in the quantum circuit composed of Haar random unitary gates and no-click measurements.
(b) Fractal dimension $D_q$ for $q=2,3$ and exponent $\tau_\textrm{Var}$ as functions of the measurement rate $p$.
}
\label{fig:IPRQuantumNoclick}
\end{figure}

In Fig.~\ref{fig:IPRQuantumNoclick}(a), we show the steady-state values of the mean IPR for $q=2$ as functions of the system size $L$.
For all measurement rates we consider, the mean IPR shows power-law scaling close to $L^{-1}$ as expected for extended trajectories. 
By performing the least-squares fitting for data with $L \geq 32$, we extract the fractal dimension $D_q$ for $q=2,3$ and the exponent $\tau_\textrm{Var}$ for the variance for the position operator. 
As shown in Fig.~\ref{fig:IPRQuantumNoclick}(b), both $D_q$ and $\tau_\textrm{Var}$ remain close to $1$ as in the absence of projective measurements. 
This result confirms the observation in Ref.~\cite{Popperl23} that forcing the measurement outcomes to be no-click ones leads to delocalization of trajectories, even though the unitary evolution is described by an Anderson chain and has a strong tendency to localization. 
Since random unitary gates in our quantum circuit rather favor delocalization, the no-click measurements with the probability $p \sim \mathcal{O}(1/L)$ are not frequent enough to stop spreading of the particle and cannot alter the delocalized nature of trajectories. 
This result also suggests irrelevance of no-click measurements for multifractality in the monitored dynamics, which could justify neglecting no-click measurements in a simplified model for a diffusive particle as discussed in Sec.~\ref{sec:StochasticResetting}.

\section{Classical system}
\label{sec:ClassicalSystem}

\subsection{Model}
\label{sec:ModelClassical}

We here present a classical analog of the monitored quantum circuit for a single particle. 
We consider a stochastic evolution of a single classical particle on a one-dimensional lattice with the length $L$. 
We then try to estimate the occupation probability $p_i$ of the particle at site $i \in \{ 1, 2, \cdots, L \}$ from limited knowledge of occupation numbers by sparse, local measurements of the particle, in addition to full knowledge of the stochastic process and initial position of the particle. 

\begin{figure}
\includegraphics[width=0.45\textwidth]{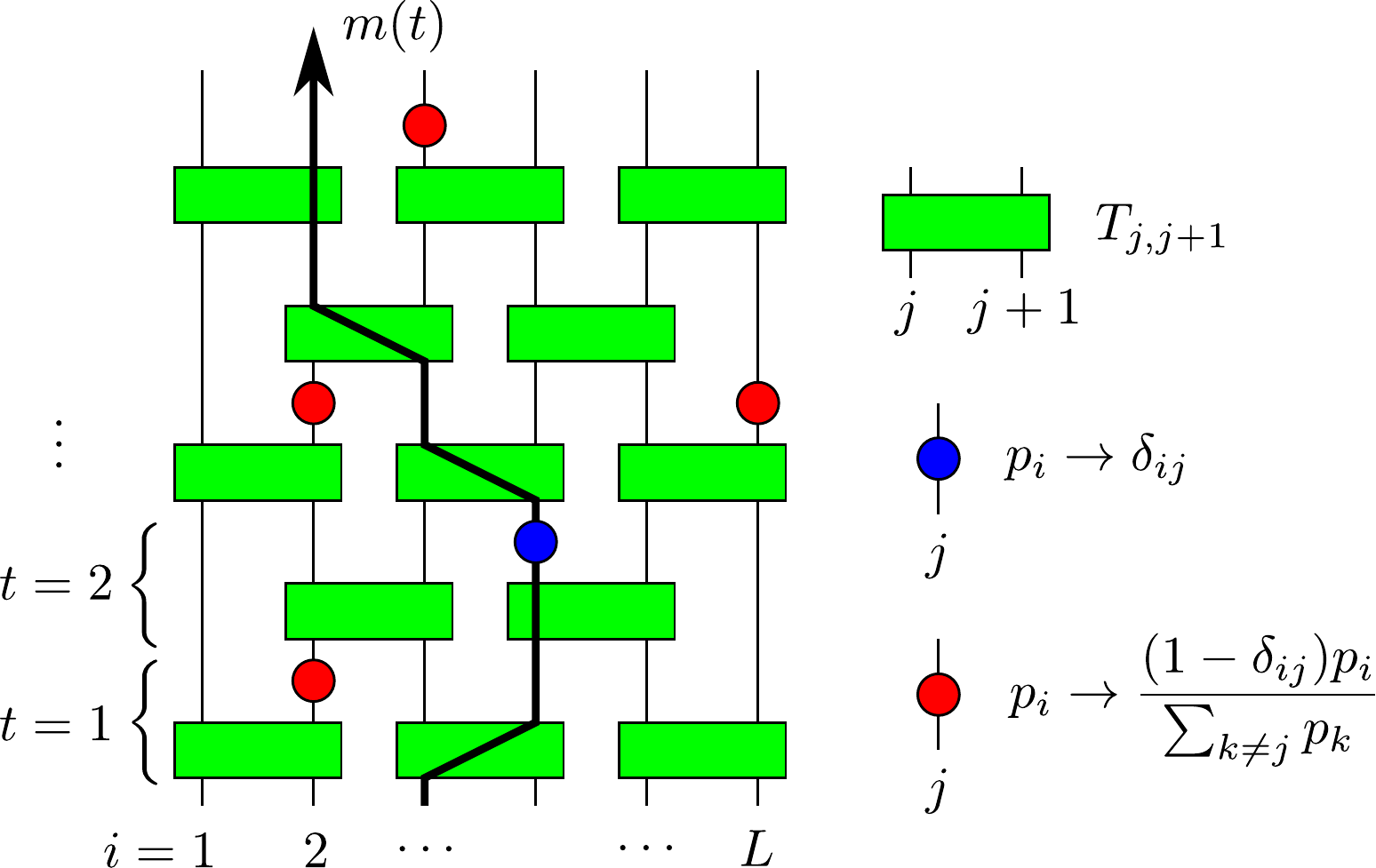}
\caption{Classical circuit consists of layers of two-site transition matrices $T_{j,j+1}$ (green rectangles). 
The horizontal and vertical direction corresponds to space and time, respectively.
The black solid line represents a trajectory $m(t)$ of the classical particle.
The particle is detected when measurement hits the particle trajectory (blue circle) whereas undetected when measurement does not hit (red circles).
}
\label{fig:ClassicalCircuit}
\end{figure}

To be more concrete, we consider the stochastic evolution of the particle by two-site transition matrices $T_{j,j+1}$ of the form, 
\begin{align} \label{eq:TransitionMatrix}
T_{j,j+1} = \begin{pmatrix} s & 1-s \\ 1-s & s \end{pmatrix},
\end{align}
where $s$ is a real parameter $s \in [0,1]$. 
The transition matrix $T_{j,j+1}$ transfers the particle from site $j$ $(j+1)$ to its neighboring site $j+1$ $(j)$ with the probability $1-s$ or leaves it at the original position with the probability $s$. 
Such transition matrices form a circuit with the brick-wall structure, similarly to the quantum case, as shown in Fig.~\ref{fig:ClassicalCircuit}. 
We always assume that the particle is initially placed at the center of the lattice, i.e., $p_i(t=0) = \delta_{i,L/2}$.
Introducing the vector notation for the occupation probabilities $\bm{p} = (p_1, \cdots, p_L)^T$, the evolution of the probability distribution $\bm{p}(t)$ by a single time step is given by 
\begin{align} \label{eq:ClassicalStochastic}
\tilde{\bm{p}}(t) = \left( \bigoplus_{j \in \mathcal{S}_t} T_{j,j+1} \right) \bm{p}(t),
\end{align}
where $\mathcal{S}_t = \{ 1,3,\cdots,L-1 \}$ for odd $t$ and $\mathcal{S}_t = \{ 2,4,\cdots,L-2 \}$ for even $t$. 
Here, we have assumed the OBC and $L$ to be even.
The real parameter $s$ for the transition matrices is either drawn randomly at each time and for each bond $(j,j+1)$ or fixed to be constant. 
If we do not measure the occupation numbers, our estimate for the probability distribution $\bm{p}(T)$ at a time $T$ is simply obtained by replacing $\tilde{\bm{p}}(t)$ by $\bm{p}(t+1)$ in Eq.~\eqref{eq:ClassicalStochastic} and repeating the evolution until $t=T$. 

We now perform measurements of the occupation number at every site with the probability $p$. 
Let $m(t) \in \{ 1,2,\cdots,L \}$ be a classical particle trajectory, which is drawn with respect to a given choice of the transition matrices $\{ T_{j,j+1} \}$ and is subject to the initial condition $m(0)=L/2$ (see also Fig.~\ref{fig:ClassicalCircuit}). 
If we are able to measure the occupation numbers of all sites (i.e., $p=1$), we have complete knowledge of the classical particle trajectory and thus obtain $p_i(t) = \delta_{i,m(t)}$. 
If the measurements are only applied to a limited number of sites, we update our estimate of the probability distribution $\bm{p}(t)$ in the following way. 
Let $A_i \equiv \{ i \}$ be a set in the sample space $\Lambda = \{ 1,2,\cdots,L \}$ and $\bar{A}_i \equiv \Lambda \backslash \{ i \}$ be its complement. 
If we denote by $X \in \Lambda$ the true position of the particle and by $Y \in \Lambda$ the position of the particle indicated by measurements, $P[Y \in A_j | X \in A_i]$ represents the conditional probability that the measurements identify the particle in $A_j$ when the particle exists in $A_i$. 
For the perfect measurement, we obviously find
\begin{align}
P[Y \in A_j | X \in A_i] &= \delta_{ij}, \\
P[Y \in \bar{A}_j | X \in A_i] &= 1-\delta_{ij}.
\end{align}
When a site $j = m(t)$ is measured and the particle is detected, the probability distribution $\tilde{\bm{p}}(t)$ is updated according to the Bayes' rule by 
\begin{align}
\tilde{p}_i(t) \to \tilde{p}'_i(t) &= \frac{\tilde{p}_i(t) P[Y \in A_j | X \in A_i]}{\sum_{k=1}^L \tilde{p}_k(t) P[Y \in A_j | X \in A_k]} \nonumber \\
&= \delta_{ij}.
\end{align}
On the other hand, when a site $j \neq m(t)$ is measured and the particle is not detected, the probability distribution is updated as 
\begin{align}
\tilde{p}_i(t) \to \tilde{p}'_i(t) &= \frac{\tilde{p}_i(t) P[Y \in \bar{A}_j | X \in A_i]}{\sum_{k=1}^L \tilde{p}_k(t) P[Y \in \bar{A}_j | X \in A_k]} \nonumber \\ 
&= \frac{(1-\delta_{ij}) \tilde{p}_i(t)}{\sum_{k \neq j} \tilde{p}_k(t)}.
\end{align}
If this measurement process scans all sites, each with the probability $p$, we obtain an estimate for the probability distribution $\bm{p}(t+1) = \tilde{\bm{p}}'(t)$ and the time evolution by a single time step is accomplished. 
Alternating layers of transition matrices and measurements are applied until $t=T$ to yield steady-state values for quantities of interest, which are computed from the estimated probability distribution $\bm{p}(T)$. 

In this classical circuit model, the classical particle trajectory $m(t)$ is solely determined by a set of the transition matrices $\{ T_{j,j+1} \}$ and is not affected by measurements. 
On the other hand, our estimate of the probability distribution $\bm{p}(t)$ can be improved by measurements of the occupation numbers at every site with the probability $p$, in combination with given information of $\{ T_{j,j+1} \}$. 
Similarly to the quantum case, the probability distribution $\bm{p}(t)$ quickly collapses into the trajectory $m(t)$ by frequent measurements with $p \sim \mathcal{O}(1)$, that is, measurements of $\mathcal{O}(L)$ sites at each time step. 
In such a case, the probability distribution $\bm{p}(t)$ is always localized. 
We thus again choose the measurement rate $p$ to scale with $1/L$, so that the particle evolved by local transition processes can escape from measurements. 
Below, we numerically demonstrate that ensembles of $\bm{p}(t)$ show multifractal scaling behaviors.

\subsection{Multifractal scaling analysis}

As in the quantum circuit model in Sec.~\ref{sec:QuantumSystem}, we study multifractality emerging in ensembles of the probability distribution $\bm{p}(t)$ estimated from measurements. 
We again consider the IPR, 
\begin{align}
\textrm{IPR}(q) \equiv \sum_{i=1}^L p_i^q, 
\end{align}
and the variance of the particle position, 
\begin{align}
\textrm{Var} = \sum_{i=1}^L i^2 p_i -\left( \sum_{i=1}^L i p_i \right)^2.
\end{align}
We are interested in the asymptotic behaviors of these quantities averaged over ensembles of the particle distribution $\bm{p}(t)$, which require three types of averaging: (i) the average over random transition matrices $T_{j,j+1}$, in case the parameter $s$ is chosen to be random, (ii) the average over measurement positions, and (iii) the average over classical trajectories $\{ m(t) \}$ drawn for a given set of $\{ T_{j,j+1} \}$. 
We denote by $\left<X\right>$ the averaged quantity of $X$. 

As discussed in Sec.~\ref{sec:Multifractal}, we consider two different averaging procedures for the IPR, the mean one $\langle \textrm{IPR}(q) \rangle$ and the typical one $e^{\langle \ln \textrm{IPR}(q) \rangle}$, for which we introduce the exponents $\tau_q$ and $\tau_q^*$, respectively, to describe their asymptotic behaviors.
We also introduce the exponent $\tau_\textrm{Var}$ for the variance of the particle position via $\langle \textrm{Var} \rangle \sim L^{2\tau_\textrm{Var}}$. 
We expect $\tau_q = \tau_q^* = q-1$ and $\tau_\textrm{Var}=1$ when the probability distributions are extended, whereas $\tau_q = \tau_q^* = \tau_\textrm{Var} = 0$ when they are localized. 
Multifractality of the probability distributions will be revealed in nonlinear functional forms of $\tau_q$ and $\tau_q^*$ in $q$. 

Along with the argument in Sec.~\ref{sec:SingleShot}, we can also consider the effect of single-shot measurements on the uniform probability distribution $p_i = 1/L$. 
In contrast to the quantum case, the uniform distribution is an exact stationary solution of the stochastic evolution process for $\bm{p}(t)$, which is described by Eq.~\eqref{eq:ClassicalStochastic} in the absence of measurements. 
In such a stationary regime, the classical particle trajectory $m(t)$ is expected to take a random value in $\{ 1,2,\cdots,L \}$. 
In this setup, the probability distribution updated according to the outcomes of single-shot measurements gives the same results for the exponents of IPRs, as given in Eqs.~\eqref{eq:TauSingleShot} and \eqref{eq:Tau*SingleShot}. 
Thus, we expect generically distinct behaviors between the exponents $\tau_q$ and $\tau_q^*$ in the presence of measurements.
As we will see below, the classical circuit model shows multifractal scaling behaviors of the IPR qualitatively similar to those for the quantum circuit model with projective measurements.

\subsection{Numerical results}
\label{sec:NumClassical}

We present our numerical results on the multifractal scaling properties of single-particle probability distributions in the classical circuit model. 
We consider a classical particle evolved either by random transition matrices (Sec.~\ref{sec:NumClassicalRandom}) or by fixed transition matrices (Sec.~\ref{sec:NumClassicalFixed}). 
In both cases, multifractality appears when the measurement outcomes are incorporated in estimating the probability distributions of the classical particle.

\subsubsection{Random transition matrix evolution}
\label{sec:NumClassicalRandom}

\begin{figure}
\includegraphics[width=0.47\textwidth]{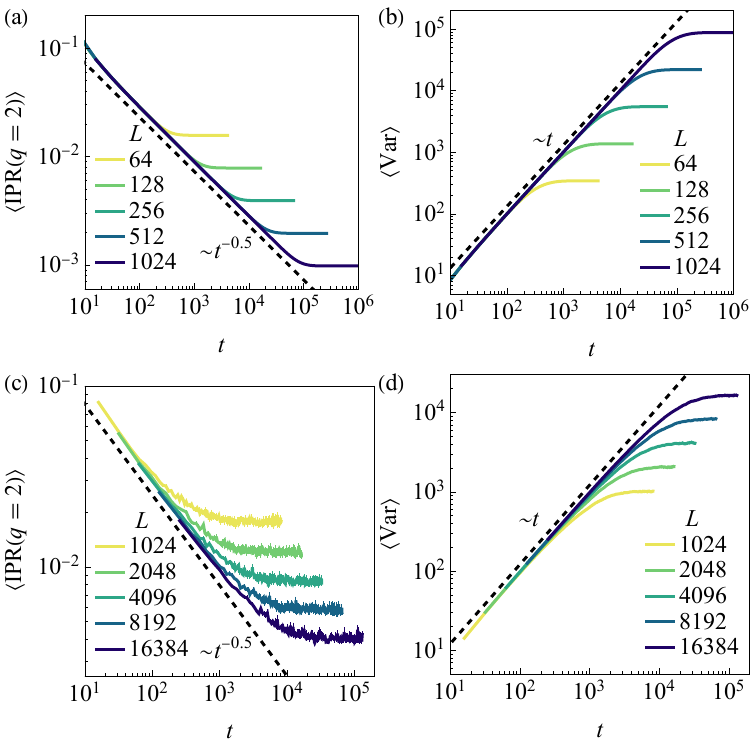}
\caption{Time evolution of the mean IPR for $q=2$ and the variance of the particle position in the classical circuit composed of random transition matrices and measurements. 
The results are shown for the measurement rate $p=0$ in (a) and (b) and for $p=1/L$ in (c) and (d).
}
\label{fig:TimeEvolClassicalRandom}
\end{figure}

We first consider the classical circuit composed of two-site transition matrices as defined in Eq.~\eqref{eq:TransitionMatrix}, in which the parameter $s$ is randomly drawn from the uniform distribution in $[0,1]$ at each time and for each bond. 
Measurements of the occupation numbers are performed at every site with the probability $p \sim \mathcal{O}(1/L)$ to improve the probability distribution $\bm{p}(t)$ for an estimate of the classical particle trajectory $m(t)$. 
In Figs.~\ref{fig:TimeEvolClassicalRandom}(a) and (b), we show time evolution of the mean IPR for $q=2$ and the variance of the particle position, respectively, in the absence of measurements, both of which are averaged over 1000 different realizations of the probability distributions. 
As in the quantum case with random unitary gates, the IPR decays as $t^{-1/2}$ and the variance grows as $t$, indicating diffusive spreading of the particle during the random transition process. 
In Figs.~\ref{fig:TimeEvolClassicalRandom}(c) and (d), we show time evolution of the same quantities in the presence of measurements with the rate $p=1/L$, which are now averaged over 8000 different realizations. 
The IPR and variance decay or grow in the same diffusive scaling form at early time for $p=1/L$ but saturate to steady-state values faster than the case for $p=0$. 
In the following analysis, we focus on those steady-state values computed at $t=L^2$ for $p=0$ and at $t=8L$ for $p>0$.

\begin{figure}
\includegraphics[width=0.47\textwidth]{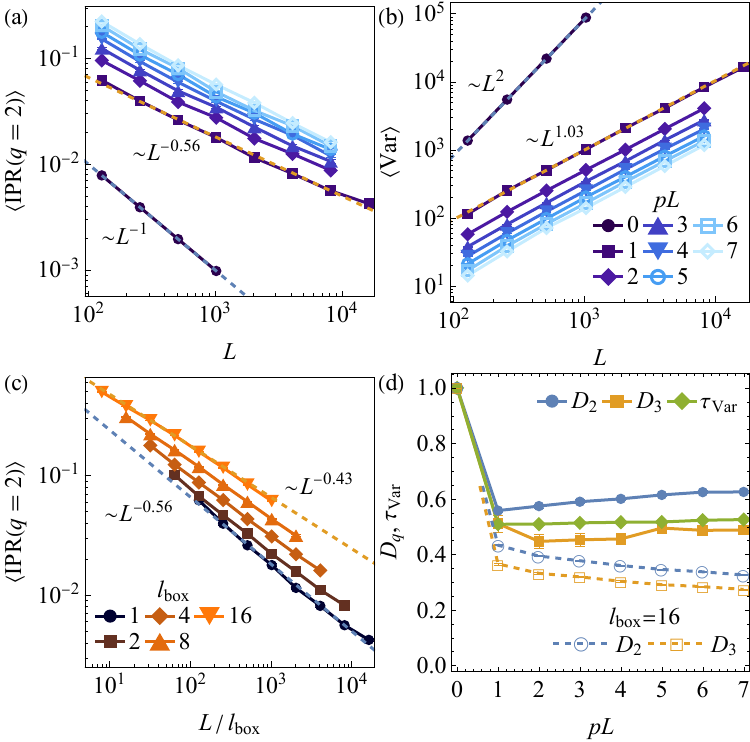}
\caption{Steady-state values of (a) the mean IPR for $q=2$ and (b) the variance of the particle position as functions of $L$ in the classical circuit composed of random transition matrices and measurements.
(c) Mean IPR for coarse-grained probability distributions with the box size $l_\textrm{box}$. 
The dashed lines are fitting functions.
(d) The fractal dimension $D_q$ for $q=2,3$ and exponent $\tau_\textrm{Var}$ are extracted from data with $L \geq 128$ and plotted against the measurement rate $p$. 
}
\label{fig:IPRClassicalRandom}
\end{figure}

The steady-state values of the mean IPR for $q=2$ and the variance of the particle position are plotted in Figs.~\ref{fig:IPRClassicalRandom}(a) and (b), respectively, as functions of the system size $L$ for several measurement rates. 
Here, the averages are taken over 1000 different realizations of the probability distribution for $p=0$ and 8000 realizations for $p>0$. 
While the steady-state probability distributions are obviously extended for $p=0$, both mean IPR and variance show power-law scaling forms with exponents $\tau_2 \simeq \tau_\textrm{Var} \simeq 0.5$ for $p=1/L$, which are clearly distinct from the extended ones. 
These exponents are also close to those found for the quantum circuit composed of random unitary gates and projective measurements in Sec.~\ref{sec:NumQuantumRandom}. 
Similarly to the quantum case, we study the instability of the multifractal exponent $\tau_q$ against coarse graining by examining the IPR for a probability distribution partitioned into boxes of the size $l_\textrm{box}$, 
\begin{align}
\mu_k(l_\textrm{box}) = \sum_{i=1}^{l_\textrm{box}} p_{i+kl_\textrm{box}}.
\end{align}
As shown in Fig.~\ref{fig:IPRClassicalRandom}(c) for $p=1/L$, the exponent $\tau_2$ varies from 0.56 ($l_\textrm{box}=1$) to 0.43 ($l_\textrm{box}=16$) by increasing the box size $l_\textrm{box}$; the latter value precisely agrees with one found in the random quantum circuit. 
In Fig.~\ref{fig:IPRClassicalRandom}(d), we show the fractal dimension $D_q$ for $q=2,3$ and the exponent $\tau_\textrm{Var}$ for the variance of the particle position, which are extracted by the least-squares fitting for data with $L \geq 128$, as functions of the measurement rate $p$. 
While $D_q$ for the coarse-grained probability distributions slowly decrease with $p$, both $D_q$ and $\tau_\textrm{Var}$ are insensitive to the variation of $p$ and take nontrivial values around 0.5 as found in the random unitary circuit with projective measurements.
These indicate that multifractality also emerges in the probability distributions for a classical particle trajectory estimated from sparse measurements. 

\begin{figure}
\includegraphics[width=0.47\textwidth]{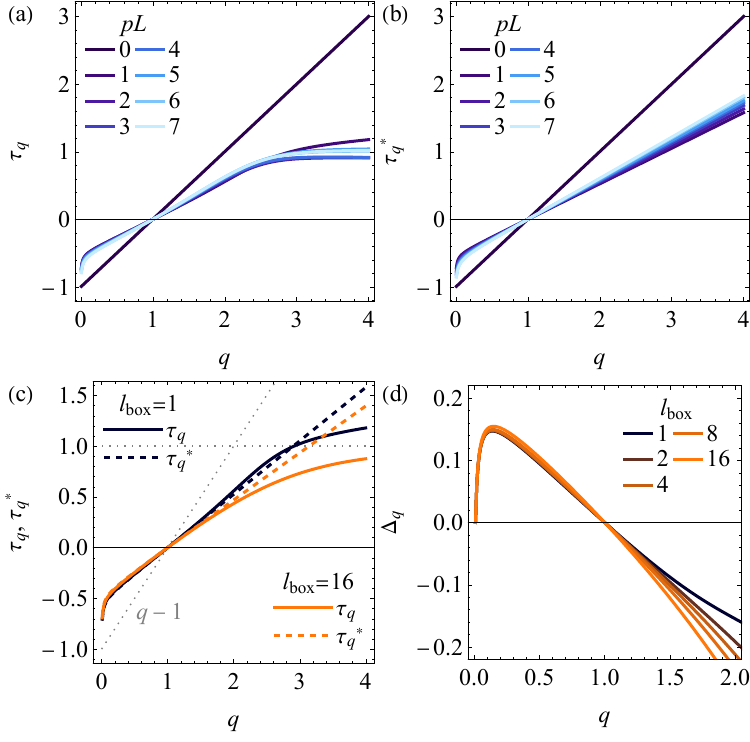}
\caption{Exponents for (a) the mean IPR and (b) the typical IPR as functions of $q$ in the circuit composed of random transition matrices and measurements. 
(c) The two exponents for $p=1/L$ are compared between the original one (black lines), coarse-grained one (orange lines), and single-shot measurement results (dotted lines). 
(d) Anomalous dimension $\Delta_q$ defined through $\tau_q = D_0(q-1)+\Delta_q$.
}
\label{fig:ExponentClassicalRandom}
\end{figure}

To study more detailed properties of multifractality, we obtain the exponent $\tau_q$ for the mean IPR and $\tau_q^*$ for the typical IPR as functions of $q$ by performing the fitting analysis for various values of $q$. 
The results are shown in Figs.~\ref{fig:ExponentClassicalRandom}(a) and (b).
In the presence of measurements, both exponents clearly deviate from the linear functional form $q-1$ for extended trajectories with strong nonlinearity around $q=0$. 
As predicted by the single-shot measurement model, the exponent $\tau_q$ for the mean IPR shows a saturating behavior towards $\tau_\infty = 1$ for $q \gtrsim 2$, which is enhanced by coarse graining [see Fig.~\ref{fig:ExponentClassicalRandom}(c)]. 
As shown in Fig.~\ref{fig:ExponentClassicalRandom}(d), the anomalous dimension $\Delta_q$ defined through $\tau_q = D_0 (q-1) +\Delta_q$ with a numerically estimated value of $D_0 \simeq 0.7$ takes a concave form, but obviously violates the symmetry about $q=1/2$ expected for critical states at the Anderson transition. 
All these features are qualitatively similar to those found in the quantum circuit with random unitary gates and projective measurements. 
A common property of the quantum circuit and the present classical circuit is diffusive spreading of the particle in the absence of measurements. 
As further substantiated in the next section, diffusive spreading is very likely to be a source of universal multifractal behaviors in monitored single-particle systems.

\subsubsection{Fixed transition matrix evolution}
\label{sec:NumClassicalFixed}

Our second example of the classical circuit consists of two-site transition matrices $T_{j,j+1}$ defined in Eq.~\eqref{eq:TransitionMatrix} with a fixed $s$. 
In particular, we here set $s=1/2$, so that the classical particle stays or hops with an equal probability by applying a transition matrix. 
In contract to the quantum circuit with fixed unitary gates where the particle spreads ballistically, this classical circuit is nothing but a discrete random walk for which diffusive spreading of the particle is naturally anticipated in the absence of measurements.
The measurements are still performed on every site with the probability $p \sim \mathcal{O}(1/L)$ to estimate the classical particle trajectory. 
Thus, in the presence of measurements, quantities of interest are averaged over (ii) measurement positions and (iii) classical particle trajectories drawn for a given set of $\{ T_{j,j+1} \}$. 
We have checked that time evolution of the mean IPR and the variance of the particle position both show similar scaling behaviors as observed in the quantum circuit with projective measurements or the classical circuit with random transition matrices.
Hereafter, we focus on the steady-state values of the IPR or variance computed at $t=L^2$ for $p=0$ and $t=8L$ for $p>0$. 
The averages are taken over 1000 trajectories for $p=0$ and 8000 trajectories for $p>0$. 

\begin{figure}
\includegraphics[width=0.47\textwidth]{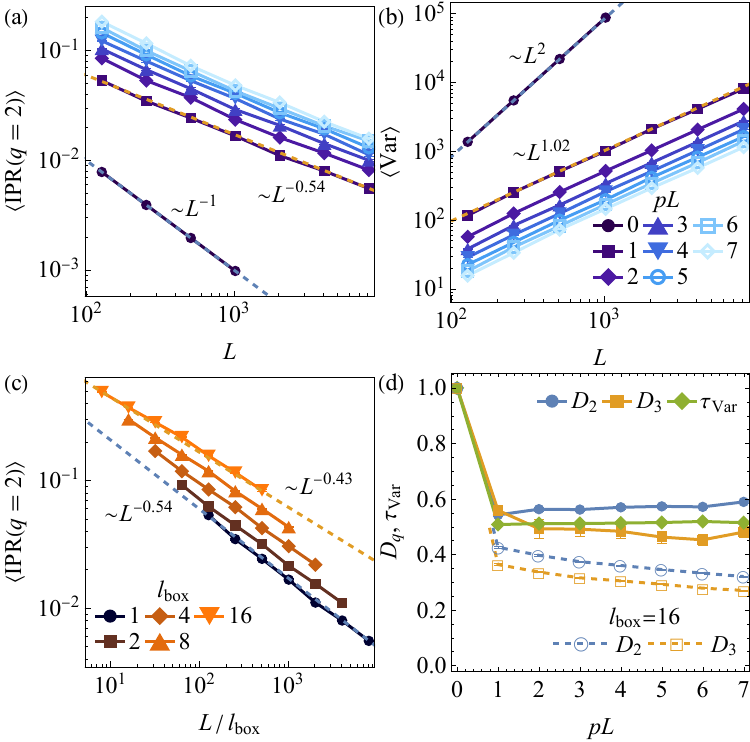}
\caption{Steady-state values of (a) the mean IPR for $q=2$ and (b) the variance of the particle position as functions of $L$ in the classical circuit composed of fixed transition matrices and measurements.
(c) Mean IPR for coarse-grained probability distributions with the box size $l_\textrm{box}$. 
The dashed lines are fitting functions.
(d) The fractal dimension $D_q$ for $q=2,3$ and exponent $\tau_\textrm{Var}$ are extracted from data with $L \geq 128$ and plotted against the measurement rate $p$. 
}
\label{fig:IPRClassicalFixed}
\end{figure}

In Figs.~\ref{fig:IPRClassicalFixed}(a) and (b), we show the steady-state values of the mean IPR for $q=2$ and the variance of the particle position, respectively, as functions of the system size $L$.
In the presence of measurements, they show power-law scaling behaviors with exponents $\tau_2 \simeq \tau_\textrm{Var} \simeq 0.5$ as already found in the other models with a diffusively spreading particle. 
As shown in Fig.~\ref{fig:IPRClassicalFixed}(c), the mean IPR for coarse-grained probability distributions has the exponent varying with the box size $l_\textrm{box}$ from 0.54 ($l_\textrm{box}=1$) to 0.43 ($l_\textrm{box}=16$); the latter value coincides with those found in the other diffusive models. 
The robustness of the fractal dimension $D_q$ and the exponent $\tau_\textrm{Var}$, which are obtained by least-squares fitting for data with $L \geq 128$, against the change of the measurement rate $p$ can also be seen from Fig.~\ref{fig:IPRClassicalFixed}(d). 

\begin{figure}
\includegraphics[width=0.47\textwidth]{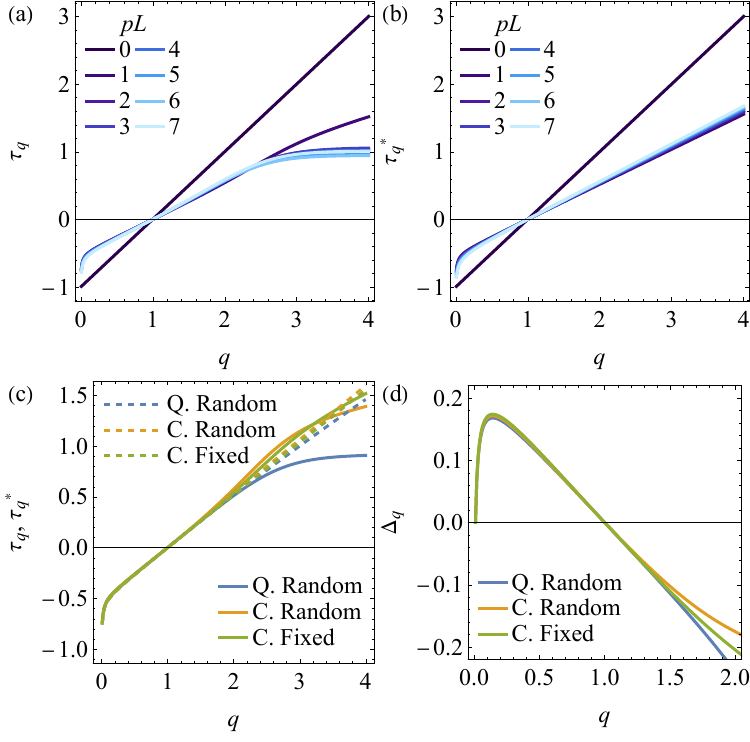}
\caption{Exponents for (a) the mean IPR and (b) the typical IPR as functions of $q$ in the circuit composed of fixed transition matrices and measurements. 
(c) The exponents $\tau_q$ (solid lines) and $\tau_q^*$ (dashed lines) are compared for $p=1/L$ between the quantum (Q) circuit with random unitary gates and the classical (C) circuit with random or fixed transition matrices. 
(d) Similar comparison made for the anomalous dimension $\Delta_q$.
}
\label{fig:ExponentClassicalFixed}
\end{figure}

We further obtain the exponent $\tau_q$ for the mean IPR and $\tau_q^*$ for the typical IPR as functions of $q$, as shown in Figs.~\ref{fig:ExponentClassicalFixed}(a) and (b), respectively.
Except that $\tau_q$ for $p=1/L$ deviates from the other data with $p > 1/L$ for $q \gtrsim 3$, possibly due to a large statistical error in the mean IPR for large $q$, they have common qualitative features with other diffusive models, such as strong nonlinearity around $q=0$, deviation from the linear functional form $q-1$ for extended trajectories, and saturating behavior of $\tau_q$ for $q \gtrsim 2$. 
In Figs.~\ref{fig:ExponentClassicalFixed}(c) and (d), we compare the multifractal exponents $\tau_q$ and $\tau_q^*$ and the anomalous dimension $\Delta_q$, which are obtained by the fitting analysis for data with $128 \leq L \leq 8192$, between three diffusive models, namely, the quantum circuit with random unitary gates and projective measurements and the classical circuits with random or fixed transition matrices and measurements. 
They show even quantitative agreement with each other for $q \lesssim 2$, implying universal scaling behaviors originated from diffusive spreading of the particle.

\subsection{Relation to random walk with stochastic resetting}
\label{sec:StochasticResetting}

From the numerical analysis, we found that the transport property of a particle in the absence of measurements, which is either diffusive or ballistic in our models, is most striking to characterize universal properties of multifractality appearing in monitored single-particle dynamics. 
Here, we argue that several multifractal properties of diffusive monitored systems are well described by a classical random walk with stochastic resetting \cite{Evans11, Evans20}. 
Specifically, we consider a random walker, which is placed on either lattice of the length $L$ or continuum, stochastically reset to the initial position at a constant rate $\lambda \sim O(1/L)$.
This involves two modifications compared with the monitored dynamics. 
First, we discard no-click measurements and only allow click measurements; a resetting event can be thought of as a collapse of the wave function or probability distribution into a localized state caused by click measurements, but there is no counterpart for no-click measurements.
Second, the measurement rate is set to be a constant, while actual measurements should occur with the probability determined by the wave function prior to the measurement (quantum case) or correlated with the probability distribution prior to the measurement (classical case). 
Despite these distinctions from the monitored dynamics, the stochastic resetting model provides qualitatively or quantitatively consistent results for multifractality in the monitored dynamics. 
We note that Ref.~\cite{Turkeshi22} has also considered stochastic resetting of quasiparticles as a phenomenological model for entanglement transitions in the quantum Ising chain subject to continuous monitoring \cite{Biella21, Turkeshi21, Piccitto22}.

\begin{figure}
\includegraphics[width=0.47\textwidth]{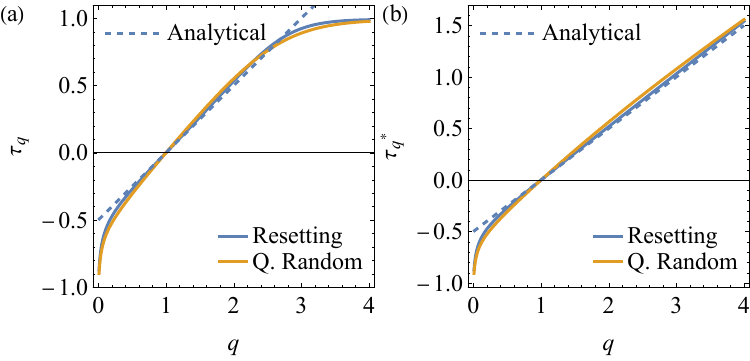}
\caption{Exponents (a) $\tau_q$ for the mean IPR and (b) $\tau_q^*$ for the typical IPR are compared between the discrete random walk subject to Poissonian resetting with the rate $\lambda=1/L$, random quantum circuit with projective measurements for $p=1/L$, and the analytical results in Eqs.~\eqref{eq:ExponentStochasticResetting1} and \eqref{eq:ExponentStochasticResetting2}. 
The numerical results are obtained by the least-squares fitting for data with $128 \leq L \leq 2048$ averaged over 8000 samples.
}
\label{fig:ExponentStochasticResetting}
\end{figure}

Let us first consider a discrete random walk subject to stochastic resetting. 
As discussed in Sec.~\ref{sec:NumClassicalFixed}, we consider a particle evolved by the brick-wall circuit composed of transition matrices with $s=1/2$ on a lattice of the length $L$ under the OBC. 
The particle is reset to the initial position $i=L/2$ by the Poissonian resetting with a rate $\lambda = 1/L$. 
After the resetting, the probability distribution becomes $p_i(t) \to \delta_{i,L/2}$.
Since time is also discrete, we draw the interval $\tau$ between two resetting events from the waiting-time distribution,
\begin{align}
\tau = \left[ -\frac{1}{\lambda} \ln \eta \right],
\end{align}
where $[x]$ is the rounding function to the nearest integer of $x$ and $\eta$ is uniformly drawn from $[0,1]$. 
We numerically generate steady-state probability distributions $\bm{p}(\infty)$ and compute the mean and typical IPR averaged over the Poissonian resetting. 
We then extract the corresponding exponents $\tau_q$ and $\tau_q^*$ by the least-squares fitting for data with $128 \leq L \leq 2048$. 
The results are shown in Fig.~\ref{fig:ExponentStochasticResetting} and further compared with the results for the quantum circuit composed of random unitary circuits and projective measurements using data with the same system sizes. 
The multifractal exponents for the stochastic resetting and monitored dynamics coincide well with each other. 

In order to analytically examine the exponents, we switch to the (infinite) continuum system where the asymptotic form of the probability distribution $p(x,t)$ for a random walker is given by the Gaussian distribution, 
\begin{align} \label{eq:GaussianDistribution}
p(x,t) = \frac{1}{\sqrt{2\pi t}} e^{-x^2/2t}.
\end{align}
The stationary distribution averaged over the Poissonian resetting with the rate $\lambda = 1/L$ is given by \cite{Evans11, Evans20}
\begin{align}
\langle p(x) \rangle = \int_0^\infty d\tau \, \lambda e^{-\lambda \tau} p(x,\tau) = \frac{1}{\sqrt{2L}} e^{\sqrt{2/L}|x|},
\end{align}
which reproduces the exponential localization of the averaged probability distribution in the monitored system (see Sec.~\ref{sec:NumQuantumRandom} and Fig.~\ref{fig:PDQuantumRandom}). 
Since the IPR for Eq.~\eqref{eq:GaussianDistribution} is obtained as 
\begin{align} \label{eq:IPRGaussian}
\textrm{IPR}(q) = \int_{-\infty}^\infty dx \, p^q(x,t) = \sqrt{\frac{(2\pi t)^{1-q}}{q}},
\end{align}
we can find the mean and typical IPR averaged over the Poissonian resetting, 
\begin{align}
\langle \textrm{IPR}(q) \rangle &= \sqrt{\frac{1}{q}\left(\frac{2\pi}{L}\right)^{1-q}} \Gamma \left( \frac{3-q}{2} \right) \ \textrm{if} \ q<3, \\
e^{\langle \ln \textrm{IPR}(q) \rangle} &= e^{\gamma (q-1)/2} \frac{1}{\sqrt{q(2\pi L)^{q-1}}},
\end{align}
where $\Gamma(x)$ is the Gamma function and $\gamma \simeq 0.577$ is the Euler's constant. 
Since the time integral of Eq.~\eqref{eq:IPRGaussian} diverges for $q \geq 3$, the mean IPR is well-behaved only for $q < 3$. 
From these expressions, we can read off the exponents, 
\begin{align}
\label{eq:ExponentStochasticResetting1}
\tau_q &= (q-1)/2 \quad (q<3), \\
\label{eq:ExponentStochasticResetting2}
\tau_q^* &= (q-1)/2.
\end{align}
As shown in Fig.~\ref{fig:ExponentStochasticResetting}, this linear functional form well coincides with the exponents $\tau_q^*$ for the typical IPR numerically obtained for the lattice models, expect for a small-$q$ region due to differences in the tails of distribution caused by boundedness of the lattice. 
The exponent $\tau_q$ for the mean IPR also coincides with the lattice models except for the small-$q$ region and for $q \gtrsim 2$. 
The latter is because the divergence of the mean IPR for $t \to 0$ can be avoided by discretization. 
This causes the deviation of $\tau_q$ from the linear functional form for $q \gtrsim 2$ in the lattice models, while the remnant of the divergence could still lead to a large statistical error for large $q$ and large $L$.
We can also evaluate the variance for the particle position by 
\begin{align}
\langle \textrm{Var} \rangle = \int_0^\infty dt \, \lambda e^{-\lambda t} \int_{-\infty}^\infty dx \, x^2 p(x,t) = L, 
\end{align}
which gives $\tau_\textrm{Var}=1/2$. 
These analytical results suggest that the monitored diffusive models have the exponent $\tau_\textrm{Var}$, as well as the fractal dimension $D_q$ for $1 \lesssim q \lesssim 2$, close to 1/2, irrespective of the measurement rate or differences in the protocols, as confirmed by the numerical results.

\section{Summary}
\label{sec:Summary}

In this work, we have shown that a single particle subject to repeated measurements exhibits multifractality revealed in the averaged IPRs conditioned on measurement outcomes. 
For the quantum systems, projective measurements give rise to multifractality of quantum trajectories and the corresponding multifractal exponents behave in qualitatively different ways depending on the diffusive or ballistic nature for the particle transport. 
However, generalized measurements with a nonzero error rate or postselection of no-click outcomes destabilize multifractality and render the trajectories extended for sufficiently large systems. 
For the classical systems, the probability distributions for random walks with improvement by measurements show multifractality similar to the diffusive quantum systems. 
We have also provided two analytically tractable models, the single-shot measurement model and random walk under Poissonian stochastic resetting, which qualitatively or even quantitatively capture several essential features of multifractality observed in the monitored systems. 

In analogy with Anderson transitions or measurement-induced phase transitions, it is natural to ask how multifractality for a monitored single particle can be generalized to higher dimensions, models with long-range hopping processes, or general graphs.
A particularly relevant question will be whether the similarity between the monitored dynamics and stochastic resetting dynamics still holds for such general situations. 
For instance, classical random walks subject to the stochastic resetting with memory of the past trajectory have been shown to exhibit an Anderson-like transition between a localized and delocalized phase \cite{Falcon-Cortes17, Boyer19}. 
It is thus interesting to investigate whether a single quantum particle evolved under adaptive dynamics with measurement and feedback undergoes such a localized-delocalized transition, which is absent in our simple monitored systems.
Another important direction will be extension to many-particle systems. 
For example, the simple exclusion process is a natural extension of the classical stochastic process from a single particle to many particles. 
It is interesting to ask whether measurements on such classical many-particle systems can lead to measurement-induced transitions reflected in the multifractal properties of quantities like the participation entropy as observed in the quantum case \cite{Sierant22}.

\section*{Acknowledgements}

We thank Akira Furusaki, Hideaki Obuse, and Toshihiro Yada for valuable discussions. 
This work was supported in part by JSPS KAKENHI Grant No. JP20K14402 and No. JP24K06897 (Y.F.)  and by JSPS KAKENHI Grant No. JP23K13037 and JST ERATO Grant No. JPMJER2302 (K.M.).
H.O. is supported by RIKEN Junior Research Associate Program. 
Part of the computation has been performed using the facilities of the Supercomputer Center, the Institute for Solid State Physics, the University of Tokyo.

\bibliography{SingleParticleMultifractality}

\begin{thebibliography}{75}%
\makeatletter
\providecommand \@ifxundefined [1]{%
 \@ifx{#1\undefined}
}%
\providecommand \@ifnum [1]{%
 \ifnum #1\expandafter \@firstoftwo
 \else \expandafter \@secondoftwo
 \fi
}%
\providecommand \@ifx [1]{%
 \ifx #1\expandafter \@firstoftwo
 \else \expandafter \@secondoftwo
 \fi
}%
\providecommand \natexlab [1]{#1}%
\providecommand \enquote  [1]{``#1''}%
\providecommand \bibnamefont  [1]{#1}%
\providecommand \bibfnamefont [1]{#1}%
\providecommand \citenamefont [1]{#1}%
\providecommand \href@noop [0]{\@secondoftwo}%
\providecommand \href [0]{\begingroup \@sanitize@url \@href}%
\providecommand \@href[1]{\@@startlink{#1}\@@href}%
\providecommand \@@href[1]{\endgroup#1\@@endlink}%
\providecommand \@sanitize@url [0]{\catcode `\\12\catcode `\$12\catcode
  `\&12\catcode `\#12\catcode `\^12\catcode `\_12\catcode `\%12\relax}%
\providecommand \@@startlink[1]{}%
\providecommand \@@endlink[0]{}%
\providecommand \url  [0]{\begingroup\@sanitize@url \@url }%
\providecommand \@url [1]{\endgroup\@href {#1}{\urlprefix }}%
\providecommand \urlprefix  [0]{URL }%
\providecommand \Eprint [0]{\href }%
\providecommand \doibase [0]{https://doi.org/}%
\providecommand \selectlanguage [0]{\@gobble}%
\providecommand \bibinfo  [0]{\@secondoftwo}%
\providecommand \bibfield  [0]{\@secondoftwo}%
\providecommand \translation [1]{[#1]}%
\providecommand \BibitemOpen [0]{}%
\providecommand \bibitemStop [0]{}%
\providecommand \bibitemNoStop [0]{.\EOS\space}%
\providecommand \EOS [0]{\spacefactor3000\relax}%
\providecommand \BibitemShut  [1]{\csname bibitem#1\endcsname}%
\let\auto@bib@innerbib\@empty
\bibitem [{\citenamefont {Evers}\ and\ \citenamefont {Mirlin}(2008)}]{Evers08}%
  \BibitemOpen
  \bibfield  {author} {\bibinfo {author} {\bibfnamefont {F.}~\bibnamefont
  {Evers}}\ and\ \bibinfo {author} {\bibfnamefont {A.~D.}\ \bibnamefont
  {Mirlin}},\ }\bibfield  {title} {\bibinfo {title} {{Anderson transitions}},\
  }\href {https://doi.org/10.1103/RevModPhys.80.1355} {\bibfield  {journal}
  {\bibinfo  {journal} {Rev. Mod. Phys.}\ }\textbf {\bibinfo {volume} {80}},\
  \bibinfo {pages} {1355} (\bibinfo {year} {2008})}\BibitemShut {NoStop}%
\bibitem [{\citenamefont {Wegner}(1980)}]{Wegner80}%
  \BibitemOpen
  \bibfield  {author} {\bibinfo {author} {\bibfnamefont {F.}~\bibnamefont
  {Wegner}},\ }\bibfield  {title} {\bibinfo {title} {{Inverse participation
  ratio in 2+$\epsilon$ dimensions}},\ }\href
  {https://doi.org/10.1007/BF01325284} {\bibfield  {journal} {\bibinfo
  {journal} {Z. Phys. B}\ }\textbf {\bibinfo {volume} {36}},\ \bibinfo {pages}
  {209} (\bibinfo {year} {1980})}\BibitemShut {NoStop}%
\bibitem [{\citenamefont {Castellani}\ and\ \citenamefont
  {Peliti}(1986)}]{Castellani86}%
  \BibitemOpen
  \bibfield  {author} {\bibinfo {author} {\bibfnamefont {C.}~\bibnamefont
  {Castellani}}\ and\ \bibinfo {author} {\bibfnamefont {L.}~\bibnamefont
  {Peliti}},\ }\bibfield  {title} {\bibinfo {title} {{Multifractal wavefunction
  at the localisation threshold}},\ }\href
  {https://doi.org/10.1088/0305-4470/19/8/004} {\bibfield  {journal} {\bibinfo
  {journal} {J. Phys. A}\ }\textbf {\bibinfo {volume} {19}},\ \bibinfo {pages}
  {L429} (\bibinfo {year} {1986})}\BibitemShut {NoStop}%
\bibitem [{\citenamefont {Janssen}(1994)}]{Janssen94}%
  \BibitemOpen
  \bibfield  {author} {\bibinfo {author} {\bibfnamefont {M.}~\bibnamefont
  {Janssen}},\ }\bibfield  {title} {\bibinfo {title} {Multifractal analysis of
  broadly-distributed observables at criticality},\ }\href
  {https://doi.org/10.1142/S021797929400049X} {\bibfield  {journal} {\bibinfo
  {journal} {Int. J. Mod. Phys. B}\ }\textbf {\bibinfo {volume} {08}},\
  \bibinfo {pages} {943} (\bibinfo {year} {1994})}\BibitemShut {NoStop}%
\bibitem [{\citenamefont {Grussbach}\ and\ \citenamefont
  {Schreiber}(1995)}]{Grussbach95}%
  \BibitemOpen
  \bibfield  {author} {\bibinfo {author} {\bibfnamefont {H.}~\bibnamefont
  {Grussbach}}\ and\ \bibinfo {author} {\bibfnamefont {M.}~\bibnamefont
  {Schreiber}},\ }\bibfield  {title} {\bibinfo {title} {{Determination of the
  mobility edge in the Anderson model of localization in three dimensions by
  multifractal analysis}},\ }\href {https://doi.org/10.1103/PhysRevB.51.663}
  {\bibfield  {journal} {\bibinfo  {journal} {Phys. Rev. B}\ }\textbf {\bibinfo
  {volume} {51}},\ \bibinfo {pages} {663} (\bibinfo {year} {1995})}\BibitemShut
  {NoStop}%
\bibitem [{\citenamefont {Mirlin}\ \emph {et~al.}(2006)\citenamefont {Mirlin},
  \citenamefont {Fyodorov}, \citenamefont {Mildenberger},\ and\ \citenamefont
  {Evers}}]{Mirlin06}%
  \BibitemOpen
  \bibfield  {author} {\bibinfo {author} {\bibfnamefont {A.~D.}\ \bibnamefont
  {Mirlin}}, \bibinfo {author} {\bibfnamefont {Y.~V.}\ \bibnamefont
  {Fyodorov}}, \bibinfo {author} {\bibfnamefont {A.}~\bibnamefont
  {Mildenberger}},\ and\ \bibinfo {author} {\bibfnamefont {F.}~\bibnamefont
  {Evers}},\ }\bibfield  {title} {\bibinfo {title} {{Exact Relations between
  Multifractal Exponents at the Anderson Transition}},\ }\href
  {https://doi.org/10.1103/PhysRevLett.97.046803} {\bibfield  {journal}
  {\bibinfo  {journal} {Phys. Rev. Lett.}\ }\textbf {\bibinfo {volume} {97}},\
  \bibinfo {pages} {046803} (\bibinfo {year} {2006})}\BibitemShut {NoStop}%
\bibitem [{\citenamefont {Subramaniam}\ \emph {et~al.}(2006)\citenamefont
  {Subramaniam}, \citenamefont {Gruzberg}, \citenamefont {Ludwig},
  \citenamefont {Evers}, \citenamefont {Mildenberger},\ and\ \citenamefont
  {Mirlin}}]{Subramaniam06}%
  \BibitemOpen
  \bibfield  {author} {\bibinfo {author} {\bibfnamefont {A.~R.}\ \bibnamefont
  {Subramaniam}}, \bibinfo {author} {\bibfnamefont {I.~A.}\ \bibnamefont
  {Gruzberg}}, \bibinfo {author} {\bibfnamefont {A.~W.~W.}\ \bibnamefont
  {Ludwig}}, \bibinfo {author} {\bibfnamefont {F.}~\bibnamefont {Evers}},
  \bibinfo {author} {\bibfnamefont {A.}~\bibnamefont {Mildenberger}},\ and\
  \bibinfo {author} {\bibfnamefont {A.~D.}\ \bibnamefont {Mirlin}},\ }\bibfield
   {title} {\bibinfo {title} {{Surface Criticality and Multifractality at
  Localization Transitions}},\ }\href
  {https://doi.org/10.1103/PhysRevLett.96.126802} {\bibfield  {journal}
  {\bibinfo  {journal} {Phys. Rev. Lett.}\ }\textbf {\bibinfo {volume} {96}},\
  \bibinfo {pages} {126802} (\bibinfo {year} {2006})}\BibitemShut {NoStop}%
\bibitem [{\citenamefont {Vasquez}\ \emph {et~al.}(2008)\citenamefont
  {Vasquez}, \citenamefont {Rodriguez},\ and\ \citenamefont
  {R\"omer}}]{Vasquez08}%
  \BibitemOpen
  \bibfield  {author} {\bibinfo {author} {\bibfnamefont {L.~J.}\ \bibnamefont
  {Vasquez}}, \bibinfo {author} {\bibfnamefont {A.}~\bibnamefont {Rodriguez}},\
  and\ \bibinfo {author} {\bibfnamefont {R.~A.}\ \bibnamefont {R\"omer}},\
  }\bibfield  {title} {\bibinfo {title} {{Multifractal analysis of the
  metal-insulator transition in the three-dimensional Anderson model. I.
  Symmetry relation under typical averaging}},\ }\href
  {https://doi.org/10.1103/PhysRevB.78.195106} {\bibfield  {journal} {\bibinfo
  {journal} {Phys. Rev. B}\ }\textbf {\bibinfo {volume} {78}},\ \bibinfo
  {pages} {195106} (\bibinfo {year} {2008})}\BibitemShut {NoStop}%
\bibitem [{\citenamefont {Rodriguez}\ \emph {et~al.}(2008)\citenamefont
  {Rodriguez}, \citenamefont {Vasquez},\ and\ \citenamefont
  {R\"omer}}]{Rodriguez08}%
  \BibitemOpen
  \bibfield  {author} {\bibinfo {author} {\bibfnamefont {A.}~\bibnamefont
  {Rodriguez}}, \bibinfo {author} {\bibfnamefont {L.~J.}\ \bibnamefont
  {Vasquez}},\ and\ \bibinfo {author} {\bibfnamefont {R.~A.}\ \bibnamefont
  {R\"omer}},\ }\bibfield  {title} {\bibinfo {title} {{Multifractal analysis of
  the metal-insulator transition in the three-dimensional Anderson model. II.
  Symmetry relation under ensemble averaging}},\ }\href
  {https://doi.org/10.1103/PhysRevB.78.195107} {\bibfield  {journal} {\bibinfo
  {journal} {Phys. Rev. B}\ }\textbf {\bibinfo {volume} {78}},\ \bibinfo
  {pages} {195107} (\bibinfo {year} {2008})}\BibitemShut {NoStop}%
\bibitem [{\citenamefont {Faez}\ \emph {et~al.}(2009)\citenamefont {Faez},
  \citenamefont {Strybulevych}, \citenamefont {Page}, \citenamefont
  {Lagendijk},\ and\ \citenamefont {van Tiggelen}}]{Faez09}%
  \BibitemOpen
  \bibfield  {author} {\bibinfo {author} {\bibfnamefont {S.}~\bibnamefont
  {Faez}}, \bibinfo {author} {\bibfnamefont {A.}~\bibnamefont {Strybulevych}},
  \bibinfo {author} {\bibfnamefont {J.~H.}\ \bibnamefont {Page}}, \bibinfo
  {author} {\bibfnamefont {A.}~\bibnamefont {Lagendijk}},\ and\ \bibinfo
  {author} {\bibfnamefont {B.~A.}\ \bibnamefont {van Tiggelen}},\ }\bibfield
  {title} {\bibinfo {title} {{Observation of Multifractality in Anderson
  Localization of Ultrasound}},\ }\href
  {https://doi.org/10.1103/PhysRevLett.103.155703} {\bibfield  {journal}
  {\bibinfo  {journal} {Phys. Rev. Lett.}\ }\textbf {\bibinfo {volume} {103}},\
  \bibinfo {pages} {155703} (\bibinfo {year} {2009})}\BibitemShut {NoStop}%
\bibitem [{\citenamefont {Obuse}\ \emph {et~al.}(2010)\citenamefont {Obuse},
  \citenamefont {Subramaniam}, \citenamefont {Furusaki}, \citenamefont
  {Gruzberg},\ and\ \citenamefont {Ludwig}}]{Obuse10}%
  \BibitemOpen
  \bibfield  {author} {\bibinfo {author} {\bibfnamefont {H.}~\bibnamefont
  {Obuse}}, \bibinfo {author} {\bibfnamefont {A.~R.}\ \bibnamefont
  {Subramaniam}}, \bibinfo {author} {\bibfnamefont {A.}~\bibnamefont
  {Furusaki}}, \bibinfo {author} {\bibfnamefont {I.~A.}\ \bibnamefont
  {Gruzberg}},\ and\ \bibinfo {author} {\bibfnamefont {A.~W.~W.}\ \bibnamefont
  {Ludwig}},\ }\bibfield  {title} {\bibinfo {title} {{Conformal invariance,
  multifractality, and finite-size scaling at Anderson localization transitions
  in two dimensions}},\ }\href {https://doi.org/10.1103/PhysRevB.82.035309}
  {\bibfield  {journal} {\bibinfo  {journal} {Phys. Rev. B}\ }\textbf {\bibinfo
  {volume} {82}},\ \bibinfo {pages} {035309} (\bibinfo {year}
  {2010})}\BibitemShut {NoStop}%
\bibitem [{\citenamefont {Rodriguez}\ \emph {et~al.}(2010)\citenamefont
  {Rodriguez}, \citenamefont {Vasquez}, \citenamefont {Slevin},\ and\
  \citenamefont {R\"omer}}]{Rodriguez10}%
  \BibitemOpen
  \bibfield  {author} {\bibinfo {author} {\bibfnamefont {A.}~\bibnamefont
  {Rodriguez}}, \bibinfo {author} {\bibfnamefont {L.~J.}\ \bibnamefont
  {Vasquez}}, \bibinfo {author} {\bibfnamefont {K.}~\bibnamefont {Slevin}},\
  and\ \bibinfo {author} {\bibfnamefont {R.~A.}\ \bibnamefont {R\"omer}},\
  }\bibfield  {title} {\bibinfo {title} {{Critical Parameters from a
  Generalized Multifractal Analysis at the Anderson Transition}},\ }\href
  {https://doi.org/10.1103/PhysRevLett.105.046403} {\bibfield  {journal}
  {\bibinfo  {journal} {Phys. Rev. Lett.}\ }\textbf {\bibinfo {volume} {105}},\
  \bibinfo {pages} {046403} (\bibinfo {year} {2010})}\BibitemShut {NoStop}%
\bibitem [{\citenamefont {Rodriguez}\ \emph {et~al.}(2011)\citenamefont
  {Rodriguez}, \citenamefont {Vasquez}, \citenamefont {Slevin},\ and\
  \citenamefont {R\"omer}}]{Rodriguez11}%
  \BibitemOpen
  \bibfield  {author} {\bibinfo {author} {\bibfnamefont {A.}~\bibnamefont
  {Rodriguez}}, \bibinfo {author} {\bibfnamefont {L.~J.}\ \bibnamefont
  {Vasquez}}, \bibinfo {author} {\bibfnamefont {K.}~\bibnamefont {Slevin}},\
  and\ \bibinfo {author} {\bibfnamefont {R.~A.}\ \bibnamefont {R\"omer}},\
  }\bibfield  {title} {\bibinfo {title} {{Multifractal finite-size scaling and
  universality at the Anderson transition}},\ }\href
  {https://doi.org/10.1103/PhysRevB.84.134209} {\bibfield  {journal} {\bibinfo
  {journal} {Phys. Rev. B}\ }\textbf {\bibinfo {volume} {84}},\ \bibinfo
  {pages} {134209} (\bibinfo {year} {2011})}\BibitemShut {NoStop}%
\bibitem [{\citenamefont {Gruzberg}\ \emph {et~al.}(2011)\citenamefont
  {Gruzberg}, \citenamefont {Ludwig}, \citenamefont {Mirlin},\ and\
  \citenamefont {Zirnbauer}}]{Gruzberg11}%
  \BibitemOpen
  \bibfield  {author} {\bibinfo {author} {\bibfnamefont {I.~A.}\ \bibnamefont
  {Gruzberg}}, \bibinfo {author} {\bibfnamefont {A.~W.~W.}\ \bibnamefont
  {Ludwig}}, \bibinfo {author} {\bibfnamefont {A.~D.}\ \bibnamefont {Mirlin}},\
  and\ \bibinfo {author} {\bibfnamefont {M.~R.}\ \bibnamefont {Zirnbauer}},\
  }\bibfield  {title} {\bibinfo {title} {{Symmetries of Multifractal Spectra
  and Field Theories of Anderson Localization}},\ }\href
  {https://doi.org/10.1103/PhysRevLett.107.086403} {\bibfield  {journal}
  {\bibinfo  {journal} {Phys. Rev. Lett.}\ }\textbf {\bibinfo {volume} {107}},\
  \bibinfo {pages} {086403} (\bibinfo {year} {2011})}\BibitemShut {NoStop}%
\bibitem [{\citenamefont {Gruzberg}\ \emph {et~al.}(2013)\citenamefont
  {Gruzberg}, \citenamefont {Mirlin},\ and\ \citenamefont
  {Zirnbauer}}]{Gruzberg13}%
  \BibitemOpen
  \bibfield  {author} {\bibinfo {author} {\bibfnamefont {I.~A.}\ \bibnamefont
  {Gruzberg}}, \bibinfo {author} {\bibfnamefont {A.~D.}\ \bibnamefont
  {Mirlin}},\ and\ \bibinfo {author} {\bibfnamefont {M.~R.}\ \bibnamefont
  {Zirnbauer}},\ }\bibfield  {title} {\bibinfo {title} {{Classification and
  symmetry properties of scaling dimensions at Anderson transitions}},\ }\href
  {https://doi.org/10.1103/PhysRevB.87.125144} {\bibfield  {journal} {\bibinfo
  {journal} {Phys. Rev. B}\ }\textbf {\bibinfo {volume} {87}},\ \bibinfo
  {pages} {125144} (\bibinfo {year} {2013})}\BibitemShut {NoStop}%
\bibitem [{\citenamefont {Padayasi}\ and\ \citenamefont
  {Gruzberg}(2023)}]{Padayasi23}%
  \BibitemOpen
  \bibfield  {author} {\bibinfo {author} {\bibfnamefont {J.}~\bibnamefont
  {Padayasi}}\ and\ \bibinfo {author} {\bibfnamefont {I.}~\bibnamefont
  {Gruzberg}},\ }\bibfield  {title} {\bibinfo {title} {{Conformal Invariance
  and Multifractality at Anderson Transitions in Arbitrary Dimensions}},\
  }\href {https://doi.org/10.1103/PhysRevLett.131.266401} {\bibfield  {journal}
  {\bibinfo  {journal} {Phys. Rev. Lett.}\ }\textbf {\bibinfo {volume} {131}},\
  \bibinfo {pages} {266401} (\bibinfo {year} {2023})}\BibitemShut {NoStop}%
\bibitem [{\citenamefont {Ludwig}\ \emph {et~al.}(1994)\citenamefont {Ludwig},
  \citenamefont {Fisher}, \citenamefont {Shankar},\ and\ \citenamefont
  {Grinstein}}]{Ludwig94}%
  \BibitemOpen
  \bibfield  {author} {\bibinfo {author} {\bibfnamefont {A.~W.~W.}\
  \bibnamefont {Ludwig}}, \bibinfo {author} {\bibfnamefont {M.~P.~A.}\
  \bibnamefont {Fisher}}, \bibinfo {author} {\bibfnamefont {R.}~\bibnamefont
  {Shankar}},\ and\ \bibinfo {author} {\bibfnamefont {G.}~\bibnamefont
  {Grinstein}},\ }\bibfield  {title} {\bibinfo {title} {{Integer quantum Hall
  transition: An alternative approach and exact results}},\ }\href
  {https://doi.org/10.1103/PhysRevB.50.7526} {\bibfield  {journal} {\bibinfo
  {journal} {Phys. Rev. B}\ }\textbf {\bibinfo {volume} {50}},\ \bibinfo
  {pages} {7526} (\bibinfo {year} {1994})}\BibitemShut {NoStop}%
\bibitem [{\citenamefont {Mudry}\ \emph {et~al.}(1996)\citenamefont {Mudry},
  \citenamefont {Chamon},\ and\ \citenamefont {Wen}}]{Mudry96}%
  \BibitemOpen
  \bibfield  {author} {\bibinfo {author} {\bibfnamefont {C.}~\bibnamefont
  {Mudry}}, \bibinfo {author} {\bibfnamefont {C.}~\bibnamefont {Chamon}},\ and\
  \bibinfo {author} {\bibfnamefont {X.-G.}\ \bibnamefont {Wen}},\ }\bibfield
  {title} {\bibinfo {title} {{Two-dimensional conformal field theory for
  disordered systems at criticality}},\ }\href
  {https://doi.org/https://doi.org/10.1016/0550-3213(96)00128-9} {\bibfield
  {journal} {\bibinfo  {journal} {Nucl. Phys. B}\ }\textbf {\bibinfo {volume}
  {466}},\ \bibinfo {pages} {383} (\bibinfo {year} {1996})}\BibitemShut
  {NoStop}%
\bibitem [{\citenamefont {Chamon}\ \emph {et~al.}(1996)\citenamefont {Chamon},
  \citenamefont {Mudry},\ and\ \citenamefont {Wen}}]{Chamon96}%
  \BibitemOpen
  \bibfield  {author} {\bibinfo {author} {\bibfnamefont {C.~d.~C.}\
  \bibnamefont {Chamon}}, \bibinfo {author} {\bibfnamefont {C.}~\bibnamefont
  {Mudry}},\ and\ \bibinfo {author} {\bibfnamefont {X.-G.}\ \bibnamefont
  {Wen}},\ }\bibfield  {title} {\bibinfo {title} {{Localization in Two
  Dimensions, Gaussian Field Theories, and Multifractality}},\ }\href
  {https://doi.org/10.1103/PhysRevLett.77.4194} {\bibfield  {journal} {\bibinfo
   {journal} {Phys. Rev. Lett.}\ }\textbf {\bibinfo {volume} {77}},\ \bibinfo
  {pages} {4194} (\bibinfo {year} {1996})}\BibitemShut {NoStop}%
\bibitem [{\citenamefont {Evers}\ \emph {et~al.}(2001)\citenamefont {Evers},
  \citenamefont {Mildenberger},\ and\ \citenamefont {Mirlin}}]{Evers01}%
  \BibitemOpen
  \bibfield  {author} {\bibinfo {author} {\bibfnamefont {F.}~\bibnamefont
  {Evers}}, \bibinfo {author} {\bibfnamefont {A.}~\bibnamefont
  {Mildenberger}},\ and\ \bibinfo {author} {\bibfnamefont {A.~D.}\ \bibnamefont
  {Mirlin}},\ }\bibfield  {title} {\bibinfo {title} {{Multifractality of wave
  functions at the quantum Hall transition revisited}},\ }\href
  {https://doi.org/10.1103/PhysRevB.64.241303} {\bibfield  {journal} {\bibinfo
  {journal} {Phys. Rev. B}\ }\textbf {\bibinfo {volume} {64}},\ \bibinfo
  {pages} {241303} (\bibinfo {year} {2001})}\BibitemShut {NoStop}%
\bibitem [{\citenamefont {Evers}\ \emph {et~al.}(2008)\citenamefont {Evers},
  \citenamefont {Mildenberger},\ and\ \citenamefont {Mirlin}}]{Evers08PRL}%
  \BibitemOpen
  \bibfield  {author} {\bibinfo {author} {\bibfnamefont {F.}~\bibnamefont
  {Evers}}, \bibinfo {author} {\bibfnamefont {A.}~\bibnamefont
  {Mildenberger}},\ and\ \bibinfo {author} {\bibfnamefont {A.~D.}\ \bibnamefont
  {Mirlin}},\ }\bibfield  {title} {\bibinfo {title} {{Multifractality at the
  Quantum Hall Transition: Beyond the Parabolic Paradigm}},\ }\href
  {https://doi.org/10.1103/PhysRevLett.101.116803} {\bibfield  {journal}
  {\bibinfo  {journal} {Phys. Rev. Lett.}\ }\textbf {\bibinfo {volume} {101}},\
  \bibinfo {pages} {116803} (\bibinfo {year} {2008})}\BibitemShut {NoStop}%
\bibitem [{\citenamefont {Obuse}\ \emph {et~al.}(2008)\citenamefont {Obuse},
  \citenamefont {Subramaniam}, \citenamefont {Furusaki}, \citenamefont
  {Gruzberg},\ and\ \citenamefont {Ludwig}}]{Obuse08}%
  \BibitemOpen
  \bibfield  {author} {\bibinfo {author} {\bibfnamefont {H.}~\bibnamefont
  {Obuse}}, \bibinfo {author} {\bibfnamefont {A.~R.}\ \bibnamefont
  {Subramaniam}}, \bibinfo {author} {\bibfnamefont {A.}~\bibnamefont
  {Furusaki}}, \bibinfo {author} {\bibfnamefont {I.~A.}\ \bibnamefont
  {Gruzberg}},\ and\ \bibinfo {author} {\bibfnamefont {A.~W.~W.}\ \bibnamefont
  {Ludwig}},\ }\bibfield  {title} {\bibinfo {title} {{Boundary Multifractality
  at the Integer Quantum Hall Plateau Transition: Implications for the Critical
  Theory}},\ }\href {https://doi.org/10.1103/PhysRevLett.101.116802} {\bibfield
   {journal} {\bibinfo  {journal} {Phys. Rev. Lett.}\ }\textbf {\bibinfo
  {volume} {101}},\ \bibinfo {pages} {116802} (\bibinfo {year}
  {2008})}\BibitemShut {NoStop}%
\bibitem [{\citenamefont {Zirnbauer}(2019)}]{Zirnbauer19}%
  \BibitemOpen
  \bibfield  {author} {\bibinfo {author} {\bibfnamefont {M.~R.}\ \bibnamefont
  {Zirnbauer}},\ }\bibfield  {title} {\bibinfo {title} {{The integer quantum
  Hall plateau transition is a current algebra after all}},\ }\href
  {https://doi.org/https://doi.org/10.1016/j.nuclphysb.2019.02.017} {\bibfield
  {journal} {\bibinfo  {journal} {Nucl. Phys. B}\ }\textbf {\bibinfo {volume}
  {941}},\ \bibinfo {pages} {458} (\bibinfo {year} {2019})}\BibitemShut
  {NoStop}%
\bibitem [{\citenamefont {Mirlin}\ \emph {et~al.}(1996)\citenamefont {Mirlin},
  \citenamefont {Fyodorov}, \citenamefont {Dittes}, \citenamefont {Quezada},\
  and\ \citenamefont {Seligman}}]{Mirlin96}%
  \BibitemOpen
  \bibfield  {author} {\bibinfo {author} {\bibfnamefont {A.~D.}\ \bibnamefont
  {Mirlin}}, \bibinfo {author} {\bibfnamefont {Y.~V.}\ \bibnamefont
  {Fyodorov}}, \bibinfo {author} {\bibfnamefont {F.-M.}\ \bibnamefont
  {Dittes}}, \bibinfo {author} {\bibfnamefont {J.}~\bibnamefont {Quezada}},\
  and\ \bibinfo {author} {\bibfnamefont {T.~H.}\ \bibnamefont {Seligman}},\
  }\bibfield  {title} {\bibinfo {title} {{Transition from localized to extended
  eigenstates in the ensemble of power-law random banded matrices}},\ }\href
  {https://doi.org/10.1103/PhysRevE.54.3221} {\bibfield  {journal} {\bibinfo
  {journal} {Phys. Rev. E}\ }\textbf {\bibinfo {volume} {54}},\ \bibinfo
  {pages} {3221} (\bibinfo {year} {1996})}\BibitemShut {NoStop}%
\bibitem [{\citenamefont {Kravtsov}\ and\ \citenamefont
  {Muttalib}(1997)}]{Kravtsov97}%
  \BibitemOpen
  \bibfield  {author} {\bibinfo {author} {\bibfnamefont {V.~E.}\ \bibnamefont
  {Kravtsov}}\ and\ \bibinfo {author} {\bibfnamefont {K.~A.}\ \bibnamefont
  {Muttalib}},\ }\bibfield  {title} {\bibinfo {title} {{New Class of Random
  Matrix Ensembles with Multifractal Eigenvectors}},\ }\href
  {https://doi.org/10.1103/PhysRevLett.79.1913} {\bibfield  {journal} {\bibinfo
   {journal} {Phys. Rev. Lett.}\ }\textbf {\bibinfo {volume} {79}},\ \bibinfo
  {pages} {1913} (\bibinfo {year} {1997})}\BibitemShut {NoStop}%
\bibitem [{\citenamefont {Evers}\ and\ \citenamefont {Mirlin}(2000)}]{Evers00}%
  \BibitemOpen
  \bibfield  {author} {\bibinfo {author} {\bibfnamefont {F.}~\bibnamefont
  {Evers}}\ and\ \bibinfo {author} {\bibfnamefont {A.~D.}\ \bibnamefont
  {Mirlin}},\ }\bibfield  {title} {\bibinfo {title} {{Fluctuations of the
  Inverse Participation Ratio at the Anderson Transition}},\ }\href
  {https://doi.org/10.1103/PhysRevLett.84.3690} {\bibfield  {journal} {\bibinfo
   {journal} {Phys. Rev. Lett.}\ }\textbf {\bibinfo {volume} {84}},\ \bibinfo
  {pages} {3690} (\bibinfo {year} {2000})}\BibitemShut {NoStop}%
\bibitem [{\citenamefont {Mirlin}\ and\ \citenamefont
  {Evers}(2000)}]{Mirlin00}%
  \BibitemOpen
  \bibfield  {author} {\bibinfo {author} {\bibfnamefont {A.~D.}\ \bibnamefont
  {Mirlin}}\ and\ \bibinfo {author} {\bibfnamefont {F.}~\bibnamefont {Evers}},\
  }\bibfield  {title} {\bibinfo {title} {{Multifractality and critical
  fluctuations at the Anderson transition}},\ }\href
  {https://doi.org/10.1103/PhysRevB.62.7920} {\bibfield  {journal} {\bibinfo
  {journal} {Phys. Rev. B}\ }\textbf {\bibinfo {volume} {62}},\ \bibinfo
  {pages} {7920} (\bibinfo {year} {2000})}\BibitemShut {NoStop}%
\bibitem [{\citenamefont {Deng}\ \emph {et~al.}(2016)\citenamefont {Deng},
  \citenamefont {Altshuler}, \citenamefont {Shlyapnikov},\ and\ \citenamefont
  {Santos}}]{Deng16}%
  \BibitemOpen
  \bibfield  {author} {\bibinfo {author} {\bibfnamefont {X.}~\bibnamefont
  {Deng}}, \bibinfo {author} {\bibfnamefont {B.~L.}\ \bibnamefont {Altshuler}},
  \bibinfo {author} {\bibfnamefont {G.~V.}\ \bibnamefont {Shlyapnikov}},\ and\
  \bibinfo {author} {\bibfnamefont {L.}~\bibnamefont {Santos}},\ }\bibfield
  {title} {\bibinfo {title} {{Quantum Levy Flights and Multifractality of
  Dipolar Excitations in a Random System}},\ }\href
  {https://doi.org/10.1103/PhysRevLett.117.020401} {\bibfield  {journal}
  {\bibinfo  {journal} {Phys. Rev. Lett.}\ }\textbf {\bibinfo {volume} {117}},\
  \bibinfo {pages} {020401} (\bibinfo {year} {2016})}\BibitemShut {NoStop}%
\bibitem [{\citenamefont {Nosov}\ \emph {et~al.}(2019)\citenamefont {Nosov},
  \citenamefont {Khaymovich},\ and\ \citenamefont {Kravtsov}}]{Nosov19}%
  \BibitemOpen
  \bibfield  {author} {\bibinfo {author} {\bibfnamefont {P.~A.}\ \bibnamefont
  {Nosov}}, \bibinfo {author} {\bibfnamefont {I.~M.}\ \bibnamefont
  {Khaymovich}},\ and\ \bibinfo {author} {\bibfnamefont {V.~E.}\ \bibnamefont
  {Kravtsov}},\ }\bibfield  {title} {\bibinfo {title} {{Correlation-induced
  localization}},\ }\href {https://doi.org/10.1103/PhysRevB.99.104203}
  {\bibfield  {journal} {\bibinfo  {journal} {Phys. Rev. B}\ }\textbf {\bibinfo
  {volume} {99}},\ \bibinfo {pages} {104203} (\bibinfo {year}
  {2019})}\BibitemShut {NoStop}%
\bibitem [{\citenamefont {Duthie}\ \emph {et~al.}(2022)\citenamefont {Duthie},
  \citenamefont {Roy},\ and\ \citenamefont {Logan}}]{Duthie22}%
  \BibitemOpen
  \bibfield  {author} {\bibinfo {author} {\bibfnamefont {A.}~\bibnamefont
  {Duthie}}, \bibinfo {author} {\bibfnamefont {S.}~\bibnamefont {Roy}},\ and\
  \bibinfo {author} {\bibfnamefont {D.~E.}\ \bibnamefont {Logan}},\ }\bibfield
  {title} {\bibinfo {title} {{Anomalous multifractality in quantum chains with
  strongly correlated disorder}},\ }\href
  {https://doi.org/10.1103/PhysRevB.106.L020201} {\bibfield  {journal}
  {\bibinfo  {journal} {Phys. Rev. B}\ }\textbf {\bibinfo {volume} {106}},\
  \bibinfo {pages} {L020201} (\bibinfo {year} {2022})}\BibitemShut {NoStop}%
\bibitem [{\citenamefont {Wang}\ \emph
  {et~al.}(2016{\natexlab{a}})\citenamefont {Wang}, \citenamefont {Liu},
  \citenamefont {Xianlong},\ and\ \citenamefont {Hu}}]{JunWang16}%
  \BibitemOpen
  \bibfield  {author} {\bibinfo {author} {\bibfnamefont {J.}~\bibnamefont
  {Wang}}, \bibinfo {author} {\bibfnamefont {X.-J.}\ \bibnamefont {Liu}},
  \bibinfo {author} {\bibfnamefont {G.}~\bibnamefont {Xianlong}},\ and\
  \bibinfo {author} {\bibfnamefont {H.}~\bibnamefont {Hu}},\ }\bibfield
  {title} {\bibinfo {title} {{Phase diagram of a non-Abelian
  Aubry-Andr\'e-Harper model with $p$-wave superfluidity}},\ }\href
  {https://doi.org/10.1103/PhysRevB.93.104504} {\bibfield  {journal} {\bibinfo
  {journal} {Phys. Rev. B}\ }\textbf {\bibinfo {volume} {93}},\ \bibinfo
  {pages} {104504} (\bibinfo {year} {2016}{\natexlab{a}})}\BibitemShut
  {NoStop}%
\bibitem [{\citenamefont {Wang}\ \emph
  {et~al.}(2016{\natexlab{b}})\citenamefont {Wang}, \citenamefont {Wang},\ and\
  \citenamefont {Chen}}]{YuchengWang16}%
  \BibitemOpen
  \bibfield  {author} {\bibinfo {author} {\bibfnamefont {Y.}~\bibnamefont
  {Wang}}, \bibinfo {author} {\bibfnamefont {Y.}~\bibnamefont {Wang}},\ and\
  \bibinfo {author} {\bibfnamefont {S.}~\bibnamefont {Chen}},\ }\bibfield
  {title} {\bibinfo {title} {{Spectral statistics, finite-size scaling and
  multifractalanalysis of quasiperiodic chain with p-wave pairing}},\ }\href
  {https://doi.org/10.1140/epjb/e2016-70473-y} {\bibfield  {journal} {\bibinfo
  {journal} {Eur. Phys. J. B}\ }\textbf {\bibinfo {volume} {89}},\ \bibinfo
  {pages} {254} (\bibinfo {year} {2016}{\natexlab{b}})}\BibitemShut {NoStop}%
\bibitem [{\citenamefont {Roy}\ \emph {et~al.}(2018)\citenamefont {Roy},
  \citenamefont {Khaymovich}, \citenamefont {Das},\ and\ \citenamefont
  {Moessner}}]{Roy18}%
  \BibitemOpen
  \bibfield  {author} {\bibinfo {author} {\bibfnamefont {S.}~\bibnamefont
  {Roy}}, \bibinfo {author} {\bibfnamefont {I.~M.}\ \bibnamefont {Khaymovich}},
  \bibinfo {author} {\bibfnamefont {A.}~\bibnamefont {Das}},\ and\ \bibinfo
  {author} {\bibfnamefont {R.}~\bibnamefont {Moessner}},\ }\bibfield  {title}
  {\bibinfo {title} {{Multifractality without fine-tuning in a Floquet
  quasiperiodic chain}},\ }\href {https://doi.org/10.21468/SciPostPhys.4.5.025}
  {\bibfield  {journal} {\bibinfo  {journal} {SciPost Phys.}\ }\textbf
  {\bibinfo {volume} {4}},\ \bibinfo {pages} {025} (\bibinfo {year}
  {2018})}\BibitemShut {NoStop}%
\bibitem [{\citenamefont {\ifmmode \check{C}\else
  \v{C}\fi{}ade\ifmmode~\check{z}\else \v{z}\fi{}}\ \emph
  {et~al.}(2019)\citenamefont {\ifmmode \check{C}\else
  \v{C}\fi{}ade\ifmmode~\check{z}\else \v{z}\fi{}}, \citenamefont {Mondaini},\
  and\ \citenamefont {Sacramento}}]{Cadez19}%
  \BibitemOpen
  \bibfield  {author} {\bibinfo {author} {\bibfnamefont {T.}~\bibnamefont
  {\ifmmode \check{C}\else \v{C}\fi{}ade\ifmmode~\check{z}\else \v{z}\fi{}}},
  \bibinfo {author} {\bibfnamefont {R.}~\bibnamefont {Mondaini}},\ and\
  \bibinfo {author} {\bibfnamefont {P.~D.}\ \bibnamefont {Sacramento}},\
  }\bibfield  {title} {\bibinfo {title} {{Edge and bulk localization of Floquet
  topological superconductors}},\ }\href
  {https://doi.org/10.1103/PhysRevB.99.014301} {\bibfield  {journal} {\bibinfo
  {journal} {Phys. Rev. B}\ }\textbf {\bibinfo {volume} {99}},\ \bibinfo
  {pages} {014301} (\bibinfo {year} {2019})}\BibitemShut {NoStop}%
\bibitem [{\citenamefont {Wang}\ \emph {et~al.}(2020)\citenamefont {Wang},
  \citenamefont {Zhang}, \citenamefont {Niu}, \citenamefont {Yu},\ and\
  \citenamefont {Liu}}]{YuchengWang20}%
  \BibitemOpen
  \bibfield  {author} {\bibinfo {author} {\bibfnamefont {Y.}~\bibnamefont
  {Wang}}, \bibinfo {author} {\bibfnamefont {L.}~\bibnamefont {Zhang}},
  \bibinfo {author} {\bibfnamefont {S.}~\bibnamefont {Niu}}, \bibinfo {author}
  {\bibfnamefont {D.}~\bibnamefont {Yu}},\ and\ \bibinfo {author}
  {\bibfnamefont {X.-J.}\ \bibnamefont {Liu}},\ }\bibfield  {title} {\bibinfo
  {title} {{Realization and Detection of Nonergodic Critical Phases in an
  Optical Raman Lattice}},\ }\href
  {https://doi.org/10.1103/PhysRevLett.125.073204} {\bibfield  {journal}
  {\bibinfo  {journal} {Phys. Rev. Lett.}\ }\textbf {\bibinfo {volume} {125}},\
  \bibinfo {pages} {073204} (\bibinfo {year} {2020})}\BibitemShut {NoStop}%
\bibitem [{\citenamefont {Gon\ifmmode~\mbox{\c{c}}\else \c{c}\fi{}alves}\ \emph
  {et~al.}(2023)\citenamefont {Gon\ifmmode~\mbox{\c{c}}\else \c{c}\fi{}alves},
  \citenamefont {Amorim}, \citenamefont {Castro},\ and\ \citenamefont
  {Ribeiro}}]{Goncalves23}%
  \BibitemOpen
  \bibfield  {author} {\bibinfo {author} {\bibfnamefont {M.}~\bibnamefont
  {Gon\ifmmode~\mbox{\c{c}}\else \c{c}\fi{}alves}}, \bibinfo {author}
  {\bibfnamefont {B.}~\bibnamefont {Amorim}}, \bibinfo {author} {\bibfnamefont
  {E.~V.}\ \bibnamefont {Castro}},\ and\ \bibinfo {author} {\bibfnamefont
  {P.}~\bibnamefont {Ribeiro}},\ }\bibfield  {title} {\bibinfo {title}
  {{Critical Phase Dualities in 1D Exactly Solvable Quasiperiodic Models}},\
  }\href {https://doi.org/10.1103/PhysRevLett.131.186303} {\bibfield  {journal}
  {\bibinfo  {journal} {Phys. Rev. Lett.}\ }\textbf {\bibinfo {volume} {131}},\
  \bibinfo {pages} {186303} (\bibinfo {year} {2023})}\BibitemShut {NoStop}%
\bibitem [{\citenamefont {Shimasaki}\ \emph {et~al.}(2024)\citenamefont
  {Shimasaki}, \citenamefont {Prichard}, \citenamefont {Kondakci},
  \citenamefont {Pagett}, \citenamefont {Bai}, \citenamefont {Dotti},
  \citenamefont {Cao}, \citenamefont {Dardia}, \citenamefont {Lu},
  \citenamefont {Grover},\ and\ \citenamefont {Weld}}]{Shimasaki24}%
  \BibitemOpen
  \bibfield  {author} {\bibinfo {author} {\bibfnamefont {T.}~\bibnamefont
  {Shimasaki}}, \bibinfo {author} {\bibfnamefont {M.}~\bibnamefont {Prichard}},
  \bibinfo {author} {\bibfnamefont {H.~E.}\ \bibnamefont {Kondakci}}, \bibinfo
  {author} {\bibfnamefont {J.~E.}\ \bibnamefont {Pagett}}, \bibinfo {author}
  {\bibfnamefont {Y.}~\bibnamefont {Bai}}, \bibinfo {author} {\bibfnamefont
  {P.}~\bibnamefont {Dotti}}, \bibinfo {author} {\bibfnamefont
  {A.}~\bibnamefont {Cao}}, \bibinfo {author} {\bibfnamefont {A.~R.}\
  \bibnamefont {Dardia}}, \bibinfo {author} {\bibfnamefont {T.-C.}\
  \bibnamefont {Lu}}, \bibinfo {author} {\bibfnamefont {T.}~\bibnamefont
  {Grover}},\ and\ \bibinfo {author} {\bibfnamefont {D.~M.}\ \bibnamefont
  {Weld}},\ }\bibfield  {title} {\bibinfo {title} {{Anomalous localization in a
  kicked quasicrystal}},\ }\href {https://doi.org/10.1038/s41567-023-02329-4}
  {\bibfield  {journal} {\bibinfo  {journal} {Nat. Phys.}\ }\textbf {\bibinfo
  {volume} {20}},\ \bibinfo {pages} {409} (\bibinfo {year} {2024})}\BibitemShut
  {NoStop}%
\bibitem [{\citenamefont {Karcher}\ \emph {et~al.}(2024)\citenamefont
  {Karcher}, \citenamefont {Vasseur},\ and\ \citenamefont
  {Gopalakrishnan}}]{Karcher24}%
  \BibitemOpen
  \bibfield  {author} {\bibinfo {author} {\bibfnamefont {J.~F.}\ \bibnamefont
  {Karcher}}, \bibinfo {author} {\bibfnamefont {R.}~\bibnamefont {Vasseur}},\
  and\ \bibinfo {author} {\bibfnamefont {S.}~\bibnamefont {Gopalakrishnan}},\
  }\bibfield  {title} {\bibinfo {title} {{Extended critical phase in
  quasiperiodic quantum Hall systems}},\ }\href
  {https://doi.org/10.1103/PhysRevB.109.064208} {\bibfield  {journal} {\bibinfo
   {journal} {Phys. Rev. B}\ }\textbf {\bibinfo {volume} {109}},\ \bibinfo
  {pages} {064208} (\bibinfo {year} {2024})}\BibitemShut {NoStop}%
\bibitem [{\citenamefont {Li}\ \emph {et~al.}(2018)\citenamefont {Li},
  \citenamefont {Chen},\ and\ \citenamefont {Fisher}}]{YaodongLi18}%
  \BibitemOpen
  \bibfield  {author} {\bibinfo {author} {\bibfnamefont {Y.}~\bibnamefont
  {Li}}, \bibinfo {author} {\bibfnamefont {X.}~\bibnamefont {Chen}},\ and\
  \bibinfo {author} {\bibfnamefont {M.~P.~A.}\ \bibnamefont {Fisher}},\
  }\bibfield  {title} {\bibinfo {title} {{Quantum Zeno effect and the many-body
  entanglement transition}},\ }\href
  {https://doi.org/10.1103/PhysRevB.98.205136} {\bibfield  {journal} {\bibinfo
  {journal} {Phys. Rev. B}\ }\textbf {\bibinfo {volume} {98}},\ \bibinfo
  {pages} {205136} (\bibinfo {year} {2018})}\BibitemShut {NoStop}%
\bibitem [{\citenamefont {Skinner}\ \emph {et~al.}(2019)\citenamefont
  {Skinner}, \citenamefont {Ruhman},\ and\ \citenamefont {Nahum}}]{Skinner19}%
  \BibitemOpen
  \bibfield  {author} {\bibinfo {author} {\bibfnamefont {B.}~\bibnamefont
  {Skinner}}, \bibinfo {author} {\bibfnamefont {J.}~\bibnamefont {Ruhman}},\
  and\ \bibinfo {author} {\bibfnamefont {A.}~\bibnamefont {Nahum}},\ }\bibfield
   {title} {\bibinfo {title} {{Measurement-Induced Phase Transitions in the
  Dynamics of Entanglement}},\ }\href
  {https://doi.org/10.1103/PhysRevX.9.031009} {\bibfield  {journal} {\bibinfo
  {journal} {Phys. Rev. X}\ }\textbf {\bibinfo {volume} {9}},\ \bibinfo {pages}
  {031009} (\bibinfo {year} {2019})}\BibitemShut {NoStop}%
\bibitem [{\citenamefont {Chan}\ \emph {et~al.}(2019)\citenamefont {Chan},
  \citenamefont {Nandkishore}, \citenamefont {Pretko},\ and\ \citenamefont
  {Smith}}]{AmosChan19}%
  \BibitemOpen
  \bibfield  {author} {\bibinfo {author} {\bibfnamefont {A.}~\bibnamefont
  {Chan}}, \bibinfo {author} {\bibfnamefont {R.~M.}\ \bibnamefont
  {Nandkishore}}, \bibinfo {author} {\bibfnamefont {M.}~\bibnamefont
  {Pretko}},\ and\ \bibinfo {author} {\bibfnamefont {G.}~\bibnamefont
  {Smith}},\ }\bibfield  {title} {\bibinfo {title} {{Unitary-projective
  entanglement dynamics}},\ }\href {https://doi.org/10.1103/PhysRevB.99.224307}
  {\bibfield  {journal} {\bibinfo  {journal} {Phys. Rev. B}\ }\textbf {\bibinfo
  {volume} {99}},\ \bibinfo {pages} {224307} (\bibinfo {year}
  {2019})}\BibitemShut {NoStop}%
\bibitem [{\citenamefont {Li}\ \emph {et~al.}(2019)\citenamefont {Li},
  \citenamefont {Chen},\ and\ \citenamefont {Fisher}}]{YaodongLi19}%
  \BibitemOpen
  \bibfield  {author} {\bibinfo {author} {\bibfnamefont {Y.}~\bibnamefont
  {Li}}, \bibinfo {author} {\bibfnamefont {X.}~\bibnamefont {Chen}},\ and\
  \bibinfo {author} {\bibfnamefont {M.~P.~A.}\ \bibnamefont {Fisher}},\
  }\bibfield  {title} {\bibinfo {title} {{Measurement-driven entanglement
  transition in hybrid quantum circuits}},\ }\href
  {https://doi.org/10.1103/PhysRevB.100.134306} {\bibfield  {journal} {\bibinfo
   {journal} {Phys. Rev. B}\ }\textbf {\bibinfo {volume} {100}},\ \bibinfo
  {pages} {134306} (\bibinfo {year} {2019})}\BibitemShut {NoStop}%
\bibitem [{\citenamefont {Szyniszewski}\ \emph {et~al.}(2019)\citenamefont
  {Szyniszewski}, \citenamefont {Romito},\ and\ \citenamefont
  {Schomerus}}]{Szyniszewski19}%
  \BibitemOpen
  \bibfield  {author} {\bibinfo {author} {\bibfnamefont {M.}~\bibnamefont
  {Szyniszewski}}, \bibinfo {author} {\bibfnamefont {A.}~\bibnamefont
  {Romito}},\ and\ \bibinfo {author} {\bibfnamefont {H.}~\bibnamefont
  {Schomerus}},\ }\bibfield  {title} {\bibinfo {title} {{Entanglement
  transition from variable-strength weak measurements}},\ }\href
  {https://doi.org/10.1103/PhysRevB.100.064204} {\bibfield  {journal} {\bibinfo
   {journal} {Phys. Rev. B}\ }\textbf {\bibinfo {volume} {100}},\ \bibinfo
  {pages} {064204} (\bibinfo {year} {2019})}\BibitemShut {NoStop}%
\bibitem [{\citenamefont {Bao}\ \emph {et~al.}(2020)\citenamefont {Bao},
  \citenamefont {Choi},\ and\ \citenamefont {Altman}}]{Bao20}%
  \BibitemOpen
  \bibfield  {author} {\bibinfo {author} {\bibfnamefont {Y.}~\bibnamefont
  {Bao}}, \bibinfo {author} {\bibfnamefont {S.}~\bibnamefont {Choi}},\ and\
  \bibinfo {author} {\bibfnamefont {E.}~\bibnamefont {Altman}},\ }\bibfield
  {title} {\bibinfo {title} {Theory of the phase transition in random unitary
  circuits with measurements},\ }\href
  {https://doi.org/10.1103/PhysRevB.101.104301} {\bibfield  {journal} {\bibinfo
   {journal} {Phys. Rev. B}\ }\textbf {\bibinfo {volume} {101}},\ \bibinfo
  {pages} {104301} (\bibinfo {year} {2020})}\BibitemShut {NoStop}%
\bibitem [{\citenamefont {Jian}\ \emph {et~al.}(2020)\citenamefont {Jian},
  \citenamefont {You}, \citenamefont {Vasseur},\ and\ \citenamefont
  {Ludwig}}]{Jian20}%
  \BibitemOpen
  \bibfield  {author} {\bibinfo {author} {\bibfnamefont {C.-M.}\ \bibnamefont
  {Jian}}, \bibinfo {author} {\bibfnamefont {Y.-Z.}\ \bibnamefont {You}},
  \bibinfo {author} {\bibfnamefont {R.}~\bibnamefont {Vasseur}},\ and\ \bibinfo
  {author} {\bibfnamefont {A.~W.~W.}\ \bibnamefont {Ludwig}},\ }\bibfield
  {title} {\bibinfo {title} {{Measurement-induced criticality in random quantum
  circuits}},\ }\href {https://doi.org/10.1103/PhysRevB.101.104302} {\bibfield
  {journal} {\bibinfo  {journal} {Phys. Rev. B}\ }\textbf {\bibinfo {volume}
  {101}},\ \bibinfo {pages} {104302} (\bibinfo {year} {2020})}\BibitemShut
  {NoStop}%
\bibitem [{\citenamefont {Gullans}\ and\ \citenamefont
  {Huse}(2020)}]{Gullans20}%
  \BibitemOpen
  \bibfield  {author} {\bibinfo {author} {\bibfnamefont {M.~J.}\ \bibnamefont
  {Gullans}}\ and\ \bibinfo {author} {\bibfnamefont {D.~A.}\ \bibnamefont
  {Huse}},\ }\bibfield  {title} {\bibinfo {title} {{Dynamical Purification
  Phase Transition Induced by Quantum Measurements}},\ }\href
  {https://doi.org/10.1103/PhysRevX.10.041020} {\bibfield  {journal} {\bibinfo
  {journal} {Phys. Rev. X}\ }\textbf {\bibinfo {volume} {10}},\ \bibinfo
  {pages} {041020} (\bibinfo {year} {2020})}\BibitemShut {NoStop}%
\bibitem [{\citenamefont {Potter}\ and\ \citenamefont
  {Vasseur}(2022)}]{Potter22}%
  \BibitemOpen
  \bibfield  {author} {\bibinfo {author} {\bibfnamefont {A.~C.}\ \bibnamefont
  {Potter}}\ and\ \bibinfo {author} {\bibfnamefont {R.}~\bibnamefont
  {Vasseur}},\ }\bibinfo {title} {{Entanglement Dynamics in Hybrid Quantum
  Circuits}},\ in\ \href {https://doi.org/10.1007/978-3-031-03998-0_9} {\emph
  {\bibinfo {booktitle} {Entanglement in Spin Chains: From Theory to Quantum
  Technology Applications}}},\ \bibinfo {editor} {edited by\ \bibinfo {editor}
  {\bibfnamefont {A.}~\bibnamefont {Bayat}}, \bibinfo {editor} {\bibfnamefont
  {S.}~\bibnamefont {Bose}},\ and\ \bibinfo {editor} {\bibfnamefont
  {H.}~\bibnamefont {Johannesson}}}\ (\bibinfo  {publisher} {Springer
  International Publishing},\ \bibinfo {address} {Cham},\ \bibinfo {year}
  {2022})\ pp.\ \bibinfo {pages} {211--249}\BibitemShut {NoStop}%
\bibitem [{\citenamefont {Lunt}\ \emph {et~al.}(2022)\citenamefont {Lunt},
  \citenamefont {Richter},\ and\ \citenamefont {Pal}}]{Lunt22}%
  \BibitemOpen
  \bibfield  {author} {\bibinfo {author} {\bibfnamefont {O.}~\bibnamefont
  {Lunt}}, \bibinfo {author} {\bibfnamefont {J.}~\bibnamefont {Richter}},\ and\
  \bibinfo {author} {\bibfnamefont {A.}~\bibnamefont {Pal}},\ }\bibinfo {title}
  {{Quantum Simulation Using Noisy Unitary Circuits and Measurements}},\ in\
  \href {https://doi.org/10.1007/978-3-031-03998-0_10} {\emph {\bibinfo
  {booktitle} {Entanglement in Spin Chains: From Theory to Quantum Technology
  Applications}}},\ \bibinfo {editor} {edited by\ \bibinfo {editor}
  {\bibfnamefont {A.}~\bibnamefont {Bayat}}, \bibinfo {editor} {\bibfnamefont
  {S.}~\bibnamefont {Bose}},\ and\ \bibinfo {editor} {\bibfnamefont
  {H.}~\bibnamefont {Johannesson}}}\ (\bibinfo  {publisher} {Springer
  International Publishing},\ \bibinfo {address} {Cham},\ \bibinfo {year}
  {2022})\ pp.\ \bibinfo {pages} {251--284}\BibitemShut {NoStop}%
\bibitem [{\citenamefont {Fisher}\ \emph {et~al.}(2023)\citenamefont {Fisher},
  \citenamefont {Khemani}, \citenamefont {Nahum},\ and\ \citenamefont
  {Vijay}}]{Fisher23}%
  \BibitemOpen
  \bibfield  {author} {\bibinfo {author} {\bibfnamefont {M.~P.}\ \bibnamefont
  {Fisher}}, \bibinfo {author} {\bibfnamefont {V.}~\bibnamefont {Khemani}},
  \bibinfo {author} {\bibfnamefont {A.}~\bibnamefont {Nahum}},\ and\ \bibinfo
  {author} {\bibfnamefont {S.}~\bibnamefont {Vijay}},\ }\bibfield  {title}
  {\bibinfo {title} {{Random Quantum Circuits}},\ }\href
  {https://doi.org/10.1146/annurev-conmatphys-031720-030658} {\bibfield
  {journal} {\bibinfo  {journal} {Ann. Rev. Condens. Matter Phys.}\ }\textbf
  {\bibinfo {volume} {14}},\ \bibinfo {pages} {335} (\bibinfo {year}
  {2023})}\BibitemShut {NoStop}%
\bibitem [{\citenamefont {Sierant}\ and\ \citenamefont
  {Turkeshi}(2022)}]{Sierant22}%
  \BibitemOpen
  \bibfield  {author} {\bibinfo {author} {\bibfnamefont {P.}~\bibnamefont
  {Sierant}}\ and\ \bibinfo {author} {\bibfnamefont {X.}~\bibnamefont
  {Turkeshi}},\ }\bibfield  {title} {\bibinfo {title} {{Universal Behavior
  beyond Multifractality of Wave Functions at Measurement-Induced Phase
  Transitions}},\ }\href {https://doi.org/10.1103/PhysRevLett.128.130605}
  {\bibfield  {journal} {\bibinfo  {journal} {Phys. Rev. Lett.}\ }\textbf
  {\bibinfo {volume} {128}},\ \bibinfo {pages} {130605} (\bibinfo {year}
  {2022})}\BibitemShut {NoStop}%
\bibitem [{\citenamefont {Iaconis}\ and\ \citenamefont
  {Chen}(2021)}]{Iaconis21}%
  \BibitemOpen
  \bibfield  {author} {\bibinfo {author} {\bibfnamefont {J.}~\bibnamefont
  {Iaconis}}\ and\ \bibinfo {author} {\bibfnamefont {X.}~\bibnamefont {Chen}},\
  }\bibfield  {title} {\bibinfo {title} {{Multifractality in nonunitary random
  dynamics}},\ }\href {https://doi.org/10.1103/PhysRevB.104.214307} {\bibfield
  {journal} {\bibinfo  {journal} {Phys. Rev. B}\ }\textbf {\bibinfo {volume}
  {104}},\ \bibinfo {pages} {214307} (\bibinfo {year} {2021})}\BibitemShut
  {NoStop}%
\bibitem [{\citenamefont {Cao}\ \emph {et~al.}(2019)\citenamefont {Cao},
  \citenamefont {Tilloy},\ and\ \citenamefont {Luca}}]{Cao19}%
  \BibitemOpen
  \bibfield  {author} {\bibinfo {author} {\bibfnamefont {X.}~\bibnamefont
  {Cao}}, \bibinfo {author} {\bibfnamefont {A.}~\bibnamefont {Tilloy}},\ and\
  \bibinfo {author} {\bibfnamefont {A.~D.}\ \bibnamefont {Luca}},\ }\bibfield
  {title} {\bibinfo {title} {{Entanglement in a fermion chain under continuous
  monitoring}},\ }\href {https://doi.org/10.21468/SciPostPhys.7.2.024}
  {\bibfield  {journal} {\bibinfo  {journal} {SciPost Phys.}\ }\textbf
  {\bibinfo {volume} {7}},\ \bibinfo {pages} {024} (\bibinfo {year}
  {2019})}\BibitemShut {NoStop}%
\bibitem [{\citenamefont {Alberton}\ \emph {et~al.}(2021)\citenamefont
  {Alberton}, \citenamefont {Buchhold},\ and\ \citenamefont
  {Diehl}}]{Alberton21}%
  \BibitemOpen
  \bibfield  {author} {\bibinfo {author} {\bibfnamefont {O.}~\bibnamefont
  {Alberton}}, \bibinfo {author} {\bibfnamefont {M.}~\bibnamefont {Buchhold}},\
  and\ \bibinfo {author} {\bibfnamefont {S.}~\bibnamefont {Diehl}},\ }\bibfield
   {title} {\bibinfo {title} {Entanglement transition in a monitored
  free-fermion chain: From extended criticality to area law},\ }\href
  {https://doi.org/10.1103/PhysRevLett.126.170602} {\bibfield  {journal}
  {\bibinfo  {journal} {Phys. Rev. Lett.}\ }\textbf {\bibinfo {volume} {126}},\
  \bibinfo {pages} {170602} (\bibinfo {year} {2021})}\BibitemShut {NoStop}%
\bibitem [{\citenamefont {Buchhold}\ \emph {et~al.}(2021)\citenamefont
  {Buchhold}, \citenamefont {Minoguchi}, \citenamefont {Altland},\ and\
  \citenamefont {Diehl}}]{Buchhold21}%
  \BibitemOpen
  \bibfield  {author} {\bibinfo {author} {\bibfnamefont {M.}~\bibnamefont
  {Buchhold}}, \bibinfo {author} {\bibfnamefont {Y.}~\bibnamefont {Minoguchi}},
  \bibinfo {author} {\bibfnamefont {A.}~\bibnamefont {Altland}},\ and\ \bibinfo
  {author} {\bibfnamefont {S.}~\bibnamefont {Diehl}},\ }\bibfield  {title}
  {\bibinfo {title} {{Effective Theory for the Measurement-Induced Phase
  Transition of Dirac Fermions}},\ }\href
  {https://doi.org/10.1103/PhysRevX.11.041004} {\bibfield  {journal} {\bibinfo
  {journal} {Phys. Rev. X}\ }\textbf {\bibinfo {volume} {11}},\ \bibinfo
  {pages} {041004} (\bibinfo {year} {2021})}\BibitemShut {NoStop}%
\bibitem [{\citenamefont {Fidkowski}\ \emph {et~al.}(2021)\citenamefont
  {Fidkowski}, \citenamefont {Haah},\ and\ \citenamefont
  {Hastings}}]{Fidkowski21}%
  \BibitemOpen
  \bibfield  {author} {\bibinfo {author} {\bibfnamefont {L.}~\bibnamefont
  {Fidkowski}}, \bibinfo {author} {\bibfnamefont {J.}~\bibnamefont {Haah}},\
  and\ \bibinfo {author} {\bibfnamefont {M.~B.}\ \bibnamefont {Hastings}},\
  }\bibfield  {title} {\bibinfo {title} {How {D}ynamical {Q}uantum {M}emories
  {F}orget},\ }\href {https://doi.org/10.22331/q-2021-01-17-382} {\bibfield
  {journal} {\bibinfo  {journal} {{Quantum}}\ }\textbf {\bibinfo {volume}
  {5}},\ \bibinfo {pages} {382} (\bibinfo {year} {2021})}\BibitemShut {NoStop}%
\bibitem [{\citenamefont {Coppola}\ \emph {et~al.}(2022)\citenamefont
  {Coppola}, \citenamefont {Tirrito}, \citenamefont {Karevski},\ and\
  \citenamefont {Collura}}]{Coppola22}%
  \BibitemOpen
  \bibfield  {author} {\bibinfo {author} {\bibfnamefont {M.}~\bibnamefont
  {Coppola}}, \bibinfo {author} {\bibfnamefont {E.}~\bibnamefont {Tirrito}},
  \bibinfo {author} {\bibfnamefont {D.}~\bibnamefont {Karevski}},\ and\
  \bibinfo {author} {\bibfnamefont {M.}~\bibnamefont {Collura}},\ }\bibfield
  {title} {\bibinfo {title} {{Growth of entanglement entropy under local
  projective measurements}},\ }\href
  {https://doi.org/10.1103/PhysRevB.105.094303} {\bibfield  {journal} {\bibinfo
   {journal} {Phys. Rev. B}\ }\textbf {\bibinfo {volume} {105}},\ \bibinfo
  {pages} {094303} (\bibinfo {year} {2022})}\BibitemShut {NoStop}%
\bibitem [{\citenamefont {Szyniszewski}\ \emph {et~al.}(2023)\citenamefont
  {Szyniszewski}, \citenamefont {Lunt},\ and\ \citenamefont
  {Pal}}]{Szyniszewski23}%
  \BibitemOpen
  \bibfield  {author} {\bibinfo {author} {\bibfnamefont {M.}~\bibnamefont
  {Szyniszewski}}, \bibinfo {author} {\bibfnamefont {O.}~\bibnamefont {Lunt}},\
  and\ \bibinfo {author} {\bibfnamefont {A.}~\bibnamefont {Pal}},\ }\bibfield
  {title} {\bibinfo {title} {{Disordered monitored free fermions}},\ }\href
  {https://doi.org/10.1103/PhysRevB.108.165126} {\bibfield  {journal} {\bibinfo
   {journal} {Phys. Rev. B}\ }\textbf {\bibinfo {volume} {108}},\ \bibinfo
  {pages} {165126} (\bibinfo {year} {2023})}\BibitemShut {NoStop}%
\bibitem [{\citenamefont {Poboiko}\ \emph {et~al.}(2023)\citenamefont
  {Poboiko}, \citenamefont {P\"opperl}, \citenamefont {Gornyi},\ and\
  \citenamefont {Mirlin}}]{Poboiko23}%
  \BibitemOpen
  \bibfield  {author} {\bibinfo {author} {\bibfnamefont {I.}~\bibnamefont
  {Poboiko}}, \bibinfo {author} {\bibfnamefont {P.}~\bibnamefont {P\"opperl}},
  \bibinfo {author} {\bibfnamefont {I.~V.}\ \bibnamefont {Gornyi}},\ and\
  \bibinfo {author} {\bibfnamefont {A.~D.}\ \bibnamefont {Mirlin}},\ }\bibfield
   {title} {\bibinfo {title} {{Theory of Free Fermions under Random Projective
  Measurements}},\ }\href {https://doi.org/10.1103/PhysRevX.13.041046}
  {\bibfield  {journal} {\bibinfo  {journal} {Phys. Rev. X}\ }\textbf {\bibinfo
  {volume} {13}},\ \bibinfo {pages} {041046} (\bibinfo {year}
  {2023})}\BibitemShut {NoStop}%
\bibitem [{\citenamefont {Chahine}\ and\ \citenamefont
  {Buchhold}(2024)}]{Chahine23}%
  \BibitemOpen
  \bibfield  {author} {\bibinfo {author} {\bibfnamefont {K.}~\bibnamefont
  {Chahine}}\ and\ \bibinfo {author} {\bibfnamefont {M.}~\bibnamefont
  {Buchhold}},\ }\bibfield  {title} {\bibinfo {title} {{Entanglement phases,
  localization, and multifractality of monitored free fermions in two
  dimensions}},\ }\href {https://doi.org/10.1103/PhysRevB.110.054313}
  {\bibfield  {journal} {\bibinfo  {journal} {Phys. Rev. B}\ }\textbf {\bibinfo
  {volume} {110}},\ \bibinfo {pages} {054313} (\bibinfo {year}
  {2024})}\BibitemShut {NoStop}%
\bibitem [{\citenamefont {Agrawal}\ \emph {et~al.}(2022)\citenamefont
  {Agrawal}, \citenamefont {Zabalo}, \citenamefont {Chen}, \citenamefont
  {Wilson}, \citenamefont {Potter}, \citenamefont {Pixley}, \citenamefont
  {Gopalakrishnan},\ and\ \citenamefont {Vasseur}}]{Agrawal22}%
  \BibitemOpen
  \bibfield  {author} {\bibinfo {author} {\bibfnamefont {U.}~\bibnamefont
  {Agrawal}}, \bibinfo {author} {\bibfnamefont {A.}~\bibnamefont {Zabalo}},
  \bibinfo {author} {\bibfnamefont {K.}~\bibnamefont {Chen}}, \bibinfo {author}
  {\bibfnamefont {J.~H.}\ \bibnamefont {Wilson}}, \bibinfo {author}
  {\bibfnamefont {A.~C.}\ \bibnamefont {Potter}}, \bibinfo {author}
  {\bibfnamefont {J.~H.}\ \bibnamefont {Pixley}}, \bibinfo {author}
  {\bibfnamefont {S.}~\bibnamefont {Gopalakrishnan}},\ and\ \bibinfo {author}
  {\bibfnamefont {R.}~\bibnamefont {Vasseur}},\ }\bibfield  {title} {\bibinfo
  {title} {{Entanglement and Charge-Sharpening Transitions in U(1) Symmetric
  Monitored Quantum Circuits}},\ }\href
  {https://doi.org/10.1103/PhysRevX.12.041002} {\bibfield  {journal} {\bibinfo
  {journal} {Phys. Rev. X}\ }\textbf {\bibinfo {volume} {12}},\ \bibinfo
  {pages} {041002} (\bibinfo {year} {2022})}\BibitemShut {NoStop}%
\bibitem [{\citenamefont {Barratt}\ \emph {et~al.}(2022)\citenamefont
  {Barratt}, \citenamefont {Agrawal}, \citenamefont {Gopalakrishnan},
  \citenamefont {Huse}, \citenamefont {Vasseur},\ and\ \citenamefont
  {Potter}}]{Barratt22}%
  \BibitemOpen
  \bibfield  {author} {\bibinfo {author} {\bibfnamefont {F.}~\bibnamefont
  {Barratt}}, \bibinfo {author} {\bibfnamefont {U.}~\bibnamefont {Agrawal}},
  \bibinfo {author} {\bibfnamefont {S.}~\bibnamefont {Gopalakrishnan}},
  \bibinfo {author} {\bibfnamefont {D.~A.}\ \bibnamefont {Huse}}, \bibinfo
  {author} {\bibfnamefont {R.}~\bibnamefont {Vasseur}},\ and\ \bibinfo {author}
  {\bibfnamefont {A.~C.}\ \bibnamefont {Potter}},\ }\bibfield  {title}
  {\bibinfo {title} {{Field Theory of Charge Sharpening in Symmetric Monitored
  Quantum Circuits}},\ }\href {https://doi.org/10.1103/PhysRevLett.129.120604}
  {\bibfield  {journal} {\bibinfo  {journal} {Phys. Rev. Lett.}\ }\textbf
  {\bibinfo {volume} {129}},\ \bibinfo {pages} {120604} (\bibinfo {year}
  {2022})}\BibitemShut {NoStop}%
\bibitem [{\citenamefont {Oshima}\ and\ \citenamefont {Fuji}(2023)}]{Oshima23}%
  \BibitemOpen
  \bibfield  {author} {\bibinfo {author} {\bibfnamefont {H.}~\bibnamefont
  {Oshima}}\ and\ \bibinfo {author} {\bibfnamefont {Y.}~\bibnamefont {Fuji}},\
  }\bibfield  {title} {\bibinfo {title} {{Charge fluctuation and
  charge-resolved entanglement in a monitored quantum circuit with $U(1)$
  symmetry}},\ }\href {https://doi.org/10.1103/PhysRevB.107.014308} {\bibfield
  {journal} {\bibinfo  {journal} {Phys. Rev. B}\ }\textbf {\bibinfo {volume}
  {107}},\ \bibinfo {pages} {014308} (\bibinfo {year} {2023})}\BibitemShut
  {NoStop}%
\bibitem [{\citenamefont {Chakraborty}\ \emph {et~al.}(2024)\citenamefont
  {Chakraborty}, \citenamefont {Chen}, \citenamefont {Zabalo}, \citenamefont
  {Wilson},\ and\ \citenamefont {Pixley}}]{Chakraborty23}%
  \BibitemOpen
  \bibfield  {author} {\bibinfo {author} {\bibfnamefont {A.}~\bibnamefont
  {Chakraborty}}, \bibinfo {author} {\bibfnamefont {K.}~\bibnamefont {Chen}},
  \bibinfo {author} {\bibfnamefont {A.}~\bibnamefont {Zabalo}}, \bibinfo
  {author} {\bibfnamefont {J.~H.}\ \bibnamefont {Wilson}},\ and\ \bibinfo
  {author} {\bibfnamefont {J.~H.}\ \bibnamefont {Pixley}},\ }\bibfield  {title}
  {\bibinfo {title} {{Charge and entanglement criticality in a U(1)-symmetric
  hybrid circuit of qubits}},\ }\href
  {https://doi.org/10.1103/PhysRevB.110.045135} {\bibfield  {journal} {\bibinfo
   {journal} {Phys. Rev. B}\ }\textbf {\bibinfo {volume} {110}},\ \bibinfo
  {pages} {045135} (\bibinfo {year} {2024})}\BibitemShut {NoStop}%
\bibitem [{\citenamefont {Jin}\ and\ \citenamefont {Martin}(2022)}]{Jin22}%
  \BibitemOpen
  \bibfield  {author} {\bibinfo {author} {\bibfnamefont {T.}~\bibnamefont
  {Jin}}\ and\ \bibinfo {author} {\bibfnamefont {D.~G.}\ \bibnamefont
  {Martin}},\ }\bibfield  {title} {\bibinfo {title} {{Kardar-Parisi-Zhang
  Physics and Phase Transition in a Classical Single Random Walker under
  Continuous Measurement}},\ }\href
  {https://doi.org/10.1103/PhysRevLett.129.260603} {\bibfield  {journal}
  {\bibinfo  {journal} {Phys. Rev. Lett.}\ }\textbf {\bibinfo {volume} {129}},\
  \bibinfo {pages} {260603} (\bibinfo {year} {2022})}\BibitemShut {NoStop}%
\bibitem [{\citenamefont {Jin}\ and\ \citenamefont {Martin}(2024)}]{Jin23}%
  \BibitemOpen
  \bibfield  {author} {\bibinfo {author} {\bibfnamefont {T.}~\bibnamefont
  {Jin}}\ and\ \bibinfo {author} {\bibfnamefont {D.~G.}\ \bibnamefont
  {Martin}},\ }\bibfield  {title} {\bibinfo {title} {Measurement-induced phase
  transition in a single-body tight-binding model},\ }\href
  {https://doi.org/10.1103/PhysRevB.110.L060202} {\bibfield  {journal}
  {\bibinfo  {journal} {Phys. Rev. B}\ }\textbf {\bibinfo {volume} {110}},\
  \bibinfo {pages} {L060202} (\bibinfo {year} {2024})}\BibitemShut {NoStop}%
\bibitem [{\citenamefont {P\"opperl}\ \emph {et~al.}(2023)\citenamefont
  {P\"opperl}, \citenamefont {Gornyi},\ and\ \citenamefont
  {Gefen}}]{Popperl23}%
  \BibitemOpen
  \bibfield  {author} {\bibinfo {author} {\bibfnamefont {P.}~\bibnamefont
  {P\"opperl}}, \bibinfo {author} {\bibfnamefont {I.~V.}\ \bibnamefont
  {Gornyi}},\ and\ \bibinfo {author} {\bibfnamefont {Y.}~\bibnamefont
  {Gefen}},\ }\bibfield  {title} {\bibinfo {title} {{Measurements on an
  Anderson chain}},\ }\href {https://doi.org/10.1103/PhysRevB.107.174203}
  {\bibfield  {journal} {\bibinfo  {journal} {Phys. Rev. B}\ }\textbf {\bibinfo
  {volume} {107}},\ \bibinfo {pages} {174203} (\bibinfo {year}
  {2023})}\BibitemShut {NoStop}%
\bibitem [{\citenamefont {Bilen}\ \emph {et~al.}(2021)\citenamefont {Bilen},
  \citenamefont {Georgeot}, \citenamefont {Giraud}, \citenamefont {Lemari\'e},\
  and\ \citenamefont {Garc\'{\i}a-Mata}}]{Bilen21}%
  \BibitemOpen
  \bibfield  {author} {\bibinfo {author} {\bibfnamefont {A.~M.}\ \bibnamefont
  {Bilen}}, \bibinfo {author} {\bibfnamefont {B.}~\bibnamefont {Georgeot}},
  \bibinfo {author} {\bibfnamefont {O.}~\bibnamefont {Giraud}}, \bibinfo
  {author} {\bibfnamefont {G.}~\bibnamefont {Lemari\'e}},\ and\ \bibinfo
  {author} {\bibfnamefont {I.}~\bibnamefont {Garc\'{\i}a-Mata}},\ }\bibfield
  {title} {\bibinfo {title} {{Symmetry violation of quantum multifractality:
  Gaussian fluctuations versus algebraic localization}},\ }\href
  {https://doi.org/10.1103/PhysRevResearch.3.L022023} {\bibfield  {journal}
  {\bibinfo  {journal} {Phys. Rev. Res.}\ }\textbf {\bibinfo {volume} {3}},\
  \bibinfo {pages} {L022023} (\bibinfo {year} {2021})}\BibitemShut {NoStop}%
\bibitem [{\citenamefont {Evans}\ and\ \citenamefont
  {Majumdar}(2011)}]{Evans11}%
  \BibitemOpen
  \bibfield  {author} {\bibinfo {author} {\bibfnamefont {M.~R.}\ \bibnamefont
  {Evans}}\ and\ \bibinfo {author} {\bibfnamefont {S.~N.}\ \bibnamefont
  {Majumdar}},\ }\bibfield  {title} {\bibinfo {title} {Diffusion with
  stochastic resetting},\ }\href
  {https://doi.org/10.1103/PhysRevLett.106.160601} {\bibfield  {journal}
  {\bibinfo  {journal} {Phys. Rev. Lett.}\ }\textbf {\bibinfo {volume} {106}},\
  \bibinfo {pages} {160601} (\bibinfo {year} {2011})}\BibitemShut {NoStop}%
\bibitem [{\citenamefont {Evans}\ \emph {et~al.}(2020)\citenamefont {Evans},
  \citenamefont {Majumdar},\ and\ \citenamefont {Schehr}}]{Evans20}%
  \BibitemOpen
  \bibfield  {author} {\bibinfo {author} {\bibfnamefont {M.~R.}\ \bibnamefont
  {Evans}}, \bibinfo {author} {\bibfnamefont {S.~N.}\ \bibnamefont
  {Majumdar}},\ and\ \bibinfo {author} {\bibfnamefont {G.}~\bibnamefont
  {Schehr}},\ }\bibfield  {title} {\bibinfo {title} {Stochastic resetting and
  applications},\ }\href {https://doi.org/10.1088/1751-8121/ab7cfe} {\bibfield
  {journal} {\bibinfo  {journal} {J. Phys. A}\ }\textbf {\bibinfo {volume}
  {53}},\ \bibinfo {pages} {193001} (\bibinfo {year} {2020})}\BibitemShut
  {NoStop}%
\bibitem [{\citenamefont {Turkeshi}\ \emph {et~al.}(2022)\citenamefont
  {Turkeshi}, \citenamefont {Dalmonte}, \citenamefont {Fazio},\ and\
  \citenamefont {Schir\`o}}]{Turkeshi22}%
  \BibitemOpen
  \bibfield  {author} {\bibinfo {author} {\bibfnamefont {X.}~\bibnamefont
  {Turkeshi}}, \bibinfo {author} {\bibfnamefont {M.}~\bibnamefont {Dalmonte}},
  \bibinfo {author} {\bibfnamefont {R.}~\bibnamefont {Fazio}},\ and\ \bibinfo
  {author} {\bibfnamefont {M.}~\bibnamefont {Schir\`o}},\ }\bibfield  {title}
  {\bibinfo {title} {{Entanglement transitions from stochastic resetting of
  non-Hermitian quasiparticles}},\ }\href
  {https://doi.org/10.1103/PhysRevB.105.L241114} {\bibfield  {journal}
  {\bibinfo  {journal} {Phys. Rev. B}\ }\textbf {\bibinfo {volume} {105}},\
  \bibinfo {pages} {L241114} (\bibinfo {year} {2022})}\BibitemShut {NoStop}%
\bibitem [{\citenamefont {Biella}\ and\ \citenamefont
  {Schir{\'{o}}}(2021)}]{Biella21}%
  \BibitemOpen
  \bibfield  {author} {\bibinfo {author} {\bibfnamefont {A.}~\bibnamefont
  {Biella}}\ and\ \bibinfo {author} {\bibfnamefont {M.}~\bibnamefont
  {Schir{\'{o}}}},\ }\bibfield  {title} {\bibinfo {title} {Many-{B}ody
  {Q}uantum {Z}eno {E}ffect and {M}easurement-{I}nduced {S}ubradiance
  {T}ransition},\ }\href {https://doi.org/10.22331/q-2021-08-19-528} {\bibfield
   {journal} {\bibinfo  {journal} {{Quantum}}\ }\textbf {\bibinfo {volume}
  {5}},\ \bibinfo {pages} {528} (\bibinfo {year} {2021})}\BibitemShut {NoStop}%
\bibitem [{\citenamefont {Turkeshi}\ \emph {et~al.}(2021)\citenamefont
  {Turkeshi}, \citenamefont {Biella}, \citenamefont {Fazio}, \citenamefont
  {Dalmonte},\ and\ \citenamefont {Schir\'o}}]{Turkeshi21}%
  \BibitemOpen
  \bibfield  {author} {\bibinfo {author} {\bibfnamefont {X.}~\bibnamefont
  {Turkeshi}}, \bibinfo {author} {\bibfnamefont {A.}~\bibnamefont {Biella}},
  \bibinfo {author} {\bibfnamefont {R.}~\bibnamefont {Fazio}}, \bibinfo
  {author} {\bibfnamefont {M.}~\bibnamefont {Dalmonte}},\ and\ \bibinfo
  {author} {\bibfnamefont {M.}~\bibnamefont {Schir\'o}},\ }\bibfield  {title}
  {\bibinfo {title} {{Measurement-induced entanglement transitions in the
  quantum Ising chain: From infinite to zero clicks}},\ }\href
  {https://doi.org/10.1103/PhysRevB.103.224210} {\bibfield  {journal} {\bibinfo
   {journal} {Phys. Rev. B}\ }\textbf {\bibinfo {volume} {103}},\ \bibinfo
  {pages} {224210} (\bibinfo {year} {2021})}\BibitemShut {NoStop}%
\bibitem [{\citenamefont {Piccitto}\ \emph {et~al.}(2022)\citenamefont
  {Piccitto}, \citenamefont {Russomanno},\ and\ \citenamefont
  {Rossini}}]{Piccitto22}%
  \BibitemOpen
  \bibfield  {author} {\bibinfo {author} {\bibfnamefont {G.}~\bibnamefont
  {Piccitto}}, \bibinfo {author} {\bibfnamefont {A.}~\bibnamefont
  {Russomanno}},\ and\ \bibinfo {author} {\bibfnamefont {D.}~\bibnamefont
  {Rossini}},\ }\bibfield  {title} {\bibinfo {title} {{Entanglement transitions
  in the quantum Ising chain: A comparison between different unravelings of the
  same Lindbladian}},\ }\href {https://doi.org/10.1103/PhysRevB.105.064305}
  {\bibfield  {journal} {\bibinfo  {journal} {Phys. Rev. B}\ }\textbf {\bibinfo
  {volume} {105}},\ \bibinfo {pages} {064305} (\bibinfo {year}
  {2022})}\BibitemShut {NoStop}%
\bibitem [{\citenamefont {Falc\'on-Cort\'es}\ \emph {et~al.}(2017)\citenamefont
  {Falc\'on-Cort\'es}, \citenamefont {Boyer}, \citenamefont {Giuggioli},\ and\
  \citenamefont {Majumdar}}]{Falcon-Cortes17}%
  \BibitemOpen
  \bibfield  {author} {\bibinfo {author} {\bibfnamefont {A.}~\bibnamefont
  {Falc\'on-Cort\'es}}, \bibinfo {author} {\bibfnamefont {D.}~\bibnamefont
  {Boyer}}, \bibinfo {author} {\bibfnamefont {L.}~\bibnamefont {Giuggioli}},\
  and\ \bibinfo {author} {\bibfnamefont {S.~N.}\ \bibnamefont {Majumdar}},\
  }\bibfield  {title} {\bibinfo {title} {{Localization Transition Induced by
  Learning in Random Searches}},\ }\href
  {https://doi.org/10.1103/PhysRevLett.119.140603} {\bibfield  {journal}
  {\bibinfo  {journal} {Phys. Rev. Lett.}\ }\textbf {\bibinfo {volume} {119}},\
  \bibinfo {pages} {140603} (\bibinfo {year} {2017})}\BibitemShut {NoStop}%
\bibitem [{\citenamefont {Boyer}\ \emph {et~al.}(2019)\citenamefont {Boyer},
  \citenamefont {Falcón-Cortés}, \citenamefont {Giuggioli},\ and\
  \citenamefont {Majumdar}}]{Boyer19}%
  \BibitemOpen
  \bibfield  {author} {\bibinfo {author} {\bibfnamefont {D.}~\bibnamefont
  {Boyer}}, \bibinfo {author} {\bibfnamefont {A.}~\bibnamefont
  {Falcón-Cortés}}, \bibinfo {author} {\bibfnamefont {L.}~\bibnamefont
  {Giuggioli}},\ and\ \bibinfo {author} {\bibfnamefont {S.~N.}\ \bibnamefont
  {Majumdar}},\ }\bibfield  {title} {\bibinfo {title} {{Anderson-like
  localization transition of random walks with resetting}},\ }\href
  {https://doi.org/10.1088/1742-5468/ab16c2} {\bibfield  {journal} {\bibinfo
  {journal} {J. Stat. Mech.}\ }\textbf {\bibinfo {volume} {2019}},\ \bibinfo
  {pages} {053204} (\bibinfo {year} {2019})}\BibitemShut {NoStop}%
\end{thebibliography}%

\end{document}